\documentclass[usenatbib,useAMS,fleqn]{mnras}

\usepackage[T1]{fontenc}
\usepackage{ae,aecompl}

\usepackage{graphicx}	
\usepackage{amsmath}	
\usepackage{amssymb}	
\usepackage{xspace}
\usepackage{xcolor}
\usepackage{aas_macros}
\usepackage{booktabs}
\usepackage{enumitem}

\title[\textsc{WARPFIELD-EMP}]{\textsc{WARPFIELD-EMP}: The Self-Consistent Prediction of Emission Lines from Evolving HII Regions in Dense Molecular Clouds}

\author[E.~W.~Pellegrini et al.]{E.~W. Pellegrini$^{1}$\thanks{eric.pellegrini@uni-heidelberg.de}, D.~Rahner$^{1}$, S. Reissl$^{1}$, S.~C.~O. Glover$^{1}$, R. S. Klessen$^{1,2}$,
\newauthor 
L. Rousseau-Nepton$^{3}$ and R. Herrera-Camus$^{4}$ \\
$^{1}$ Universit{\"a}t Heidelberg, Zentrum f{\"u}r Astronomie, Institut f{\"u}r Theoretische Astrophysik, \\
Albert-Ueberle-Stra{\ss}e 2, 69120 Heidelberg, Germany\\
$^{2}$ Universit{\"a}t Heidelberg, Interdisziplin{\"a}res Zentrum f{\"u}r Wissenschaftliches Rechnen,\\
Im Neuenheimer Feld 205, 69120 Heidelberg, Germany \\
$^{3}$ Canada-France-Hawaii Telescope, Kamuela, HI, 96743, USA\\
$^{4}$ Departamento de Astronom\'ia, Facultad Ciencias F\'isicas y Matem\'aticas, Universidad de Concepci\'on, \\
Av.\ Esteban Iturra s/n Barrio Universitario, Casilla 160-C, Concepci\'on, Chile
}

\date{Accepted XXX; Received YYY; in original form ZZZ}

\pubyear{2019}

\newcommand*\diff{\mathop{}\!\mathrm{d}}

\newcommand{\SFE}{$\epsilon_{\rm SF}$\xspace}

\newcommand{\ccm}{cm$^{-3}$\xspace}

\newcommand{\hii}{H$\,${\sc ii}\xspace}

\newcommand{\Ha}{H$\alpha$\xspace}
\newcommand{\Hb}{H$\beta$\xspace}
\newcommand{\Hamath}{{\rm H}\alpha\xspace}
\newcommand{\Hbmath}{{\rm H}\beta\xspace}

\newcommand{\CII}{[\ion{C}{II}]\xspace}
\newcommand{\CI}{[\ion{C}{I}]\xspace}
\newcommand{\NII}{[\ion{N}{II}]\xspace}

\newcommand{\SII}{[\ion{S}{II}]\xspace}
\newcommand{\SIII}{[\ion{S}{III}]\xspace}
\newcommand{\SIV}{[\ion{S}{IV}]\xspace}
\newcommand{\SiII}{[\ion{Si}{II}]\xspace}

\newcommand{\OI}{[\ion{O}{I}]\xspace}
\newcommand{\OII}{[\ion{O}{II}]\xspace}
\newcommand{\OIII}{[\ion{O}{III}]\xspace}

\newcommand{\NeV}{[\ion{N}{V}]\xspace}
\newcommand{\NeIII}{[\ion{N}{III}]\xspace}
\newcommand{\NeII}{[\ion{Ne}{II}]\xspace}
\newcommand{\ArII}{[\ion{Ar}{II}]\xspace}
\newcommand{\ArIII}{[\ion{Ar}{III}]\xspace}

\newcommand{\twc}{$^{12}$CO\xspace}

\begin{document}

\maketitle

\begin{abstract}
We present the {\sc warpfield} emission predictor, {\sc warpfield-emp}, which couples the 1D stellar feedback code {\sc warpfield} with the {\sc cloudy} \hii region/PDR code and the {\sc polaris} radiative transfer code, in order to make detailed predictions for the time-dependent line and continuum emission arising from the H{\sc ii} region and PDR surrounding an evolving star cluster. {\sc warpfield-emp} accounts for a wide range of physical processes (stellar winds, supernovae, radiation pressure, gravity, thermal conduction, radiative cooling, dust extinction etc.) and yet runs quickly enough to allow us to explore broad ranges of different model parameters. We compare the results of an extensive set of models with SITELLE observations of a large sample of \hii regions in NGC~628 and find very good agreement, particularly for the highest signal-to-noise observations. We show that our approach of modeling individual clouds from first principles (instead of in terms of dimensionless quantities such as the ionization parameter) allows us to avoid long-standing degeneracies in the interpretation of \hii region diagnostics and enables us to relate these diagnostics to important physical parameters such as cloud mass or cluster age. Finally, we explore the implications of our models regarding the reliability of simple metallicity diagnostics, the properties of long-lived embedded clusters, and the role played by winds and supernovae in regulating \hii region and PDR line emission.
\end{abstract}

\section{Introduction}
Massive stars play a central role in the evolution of the interstellar medium (ISM) and the self-regulation of star formation. Feedback from these stars in the form of strong stellar winds, ionizing and non-ionizing radiation, and  supernova explosions injects a large quantity of energy and momentum into the ISM and is one of the main processes that regulates the star formation rate on scales comparable to individual dense molecular clouds (see e.g.~the reviews by \citealt{Zinnecker2007}, \citealt{Krumholz2014} and \citealt{Klessen2016}). In addition, in all but the very nearest star-forming systems, observations of ``star formation'' are actually observations of young massive stars, either directly (e.g.\ in the far ultraviolet continuum) or more commonly indirectly, in the form of reprocessed radiation from gas (e.g.\ \Ha, \CII) or dust (e.g.\ 24$\mu$m or far-infrared emission).
Observations of line emission from the \hii regions and photodissociation regions (PDRs) created by radiation from massive stars are also widely used as diagnostics to estimate gas properties such as density or metallicity, particularly in extragalactic systems \citep[see e.g.][]{Kewley2001,Kauffmann2003,Kewley2013,Cormier2015,Sanchez2015}.

Our understanding of stellar feedback, the ISM and how the latter responds to the former is based in large part on observations of nebular and molecular lines as well as dust emission powered by massive stars. These provide information about how radiation propagates in star-forming regions, the distribution of molecular gas, and the local heating of dust by starlight. Line intensities place constraints on gas density and emissivity, while line profiles inform us about line-of-sight gas motions, including turbulence, and large scale motion. Correctly interpreting the information provided by these various types of observations is of central importance to our understanding of star formation. 

There is therefore a great need for models capable of predicting the properties of a wide range of observational tracers as a function of the gas distribution within a star-forming cloud and the stellar population forming inside it. Unfortunately, this is a very difficult problem. One would ideally use high-resolution 3D simulations of massive star formation and feedback within molecular clouds as a basis for making these observational predictions. However, models of this type that include all of the relevant physics (photo-ionization, photo-dissociation, radiation pressure, winds and outflows, as well as supernovae) remain thin on the ground, with most 3D simulations only including some subset of these processes \citep[see e.g.][for some recent examples]{Dale2014,Rosen2016,Peters2017,Haid2018}. Moreover, these models are computationally expensive, often limiting them to relatively low resolution, and making it completely impractical to use them to explore a large parameter space. 

For this reason, the interpretation of H{\sc ii} region and PDR diagnostics has often been carried out using models that do not attempt to self-consistently follow the growth of an H{\sc ii} region and PDR around an assembling cluster of massive stars, but instead simply consider a broad grid of values for the gas density, radiation field strength and/or ionization parameter\footnote{The ionization parameter ${\cal U}$ is the ratio of the number density of ionizing photons to the number density of hydrogen atoms, i.e.\ ${\cal U} = n_{\gamma, {\rm ion}}/{n_{\rm H}}$.}, the composite spectrum of the stellar cluster, and often also the metallicity \citep[see e.g.][for some influential examples of this type of model]{Kaufman1999,Kewley2001,Kewley2002}. More recently the calculations have also systematically included variable cluster ages and time-varying spectral energy distributions (SEDs) \citep[see][]{Byler2017}, while not solving for the coupling between cloud and cluster age.
This approach is convenient and computationally efficient, but relies on the often unstated assumption that these parameters are not strongly correlated, such that it is meaningful to vary them independently of each other. As we will see later, this is an over-simplification that does not correctly describe the behaviour of real H{\sc ii} regions. 

One-dimensional dynamical models offer a useful compromise between computationally expensive 3D models and computationally efficient but physically unrealistic static models. Although these models have a long history \citep[see e.g.][]{Castor1975,Weaver1977}, most make simplifying assumptions such as the neglect of gravity and/or radiation pressure that limit their applicability. In \citet[][hereafter Paper I]{Rahner2017a}, we introduced {\sc warpfield}, a new 1D code for simulating the time evolution and structure of the stellar wind bubble, H{\sc ii} region and PDR surrounding a cluster of massive stars. This code accounts self-consistently for the physics of stellar winds, supernovae, radiation pressure, ionization and gravity, allowing it to be applied to clouds of finite size that are not necessarily immediately disrupted by stellar feedback. It solves explicitly for the density structure adopted by the gas in response to the action of the feedback, with the help of the reasonable assumptions that the internal pressure of the feedback-blown bubble is much larger than the external pressure and that the shell surrounding the bubble is in quasi-hydrostatic equilibrium \citep{Abel_2005, Pellegrini2007,Pellegrini2011}. Finally, the time-dependent approach adopted in {\sc warpfield} also allows one to account for the evolution of the luminosity and composite spectrum of the stellar cluster as it ages. 

Although {\sc warpfield} has proved to be a useful and flexible code for analyzing the evolution of feedback-affected regions around star clusters \citep[see e.g.][for some initial applications]{Rahner2018,Rugel2018}, it does not by itself allow us to predict the properties of observational tracers of these regions. In the present paper, we address this by coupling {\sc warpfield} with state-of-the-art treatments of the microphysics of gas in H{\sc ii} regions and PDRs ({\sc cloudy}; see \citealt{Ferland2017}), and of line and continuum radiative transfer (RT) ({\sc polaris}; see \citealt{Reissl2016}). This coupled model, which we refer to as the {\sc warpfield} emission predictor, or {\sc warpfield-emp} for short, allows us to make detailed predictions for the time-dependent line and continuum emission arising from the H{\sc ii} region and PDR surrounding an evolving star cluster (potentially including on-going star formation). It accounts for both nebular and dust emission and also for the important effects of dust absorption on the emergent spectrum.

The structure of our paper is as follows. In Section~\ref{sec:EMP}, we review the basic features of {\sc warpfield} and describe several recent updates we have made to the code. We then discuss how we couple {\sc warpfield} to {\sc cloudy} and {\sc polaris} to construct the combined {\sc warpfield-emp} framework. We also discuss how we model the SED from the central stellar cluster. In Section~\ref{sec:results1}, we present an important application of this framework, an exploration of how the location of a star-forming region in a line diagnostic diagram such as the BPT diagram \citep{Baldwin1981} varies as the region evolves. Section~\ref{sec:obscomp} compares {\sc warpfield-emp} output with real observational data and also discusses some of the complications involved in doing so in a meaningful fashion. Section~\ref{sec:analysis} examines whether we can use the location of our model star-forming regions in the BPT diagram to draw inferences about the initial and current properties of the host cloud and stellar cluster. In Sections~\ref{sec:simple_diagnostics} and \ref{sec:embedded_objects} we discuss some important implications of our results regarding the applicability of simple metal abundance estimators and the nature of embedded clusters. In Section~\ref{sec:nature}, we explain why our dynamical approach to modelling line emission from star-forming regions is a significant improvement over previous approaches. Finally, in Section~\ref{sec:summary} we present a summary of our main results.

\section{WARPFIELD-EMP}\label{sec:EMP}

\begin{figure*}
	\begin{center} 
		\includegraphics[width=0.315\textwidth]{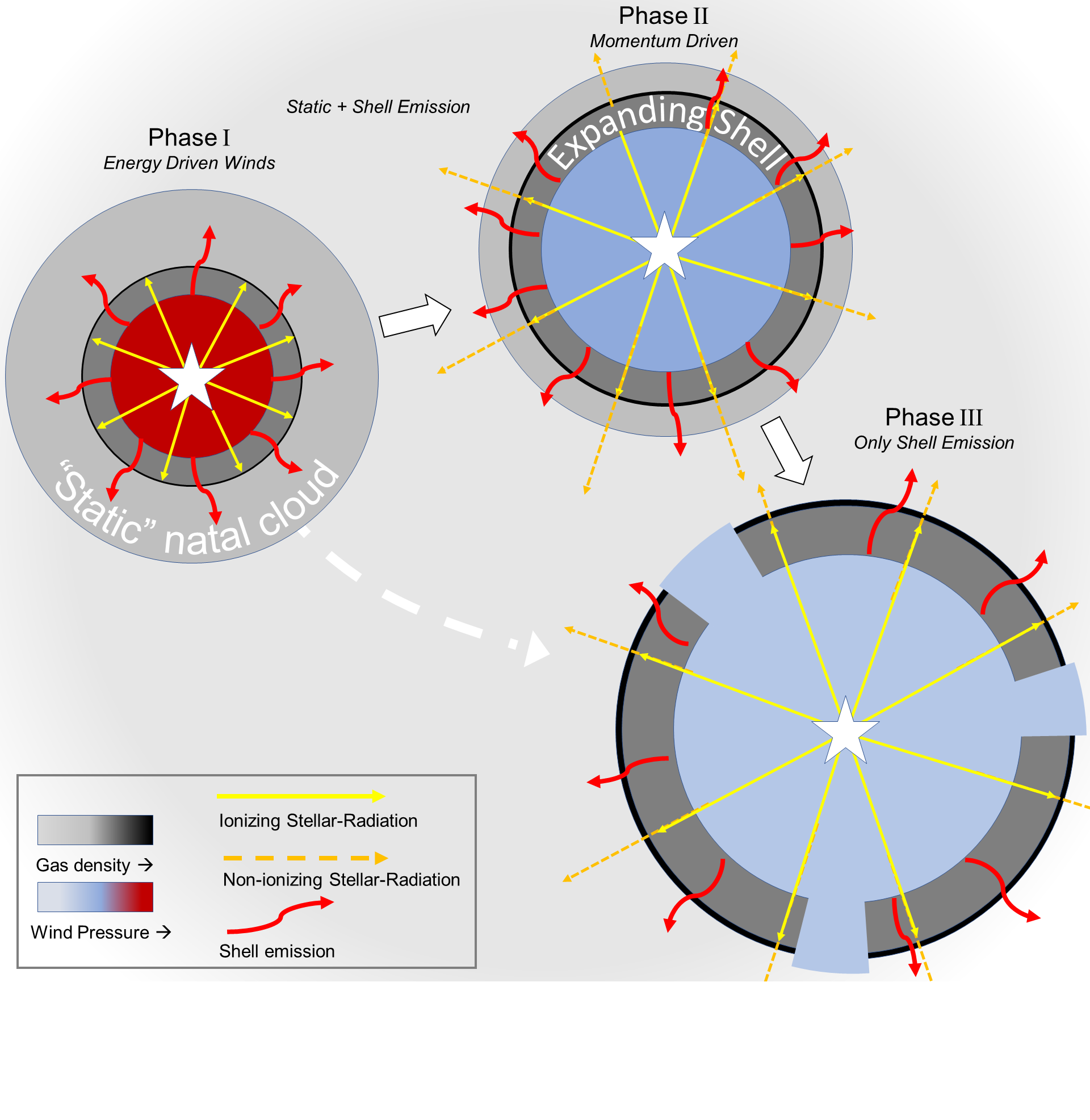}
		\includegraphics[width=0.60\textwidth]{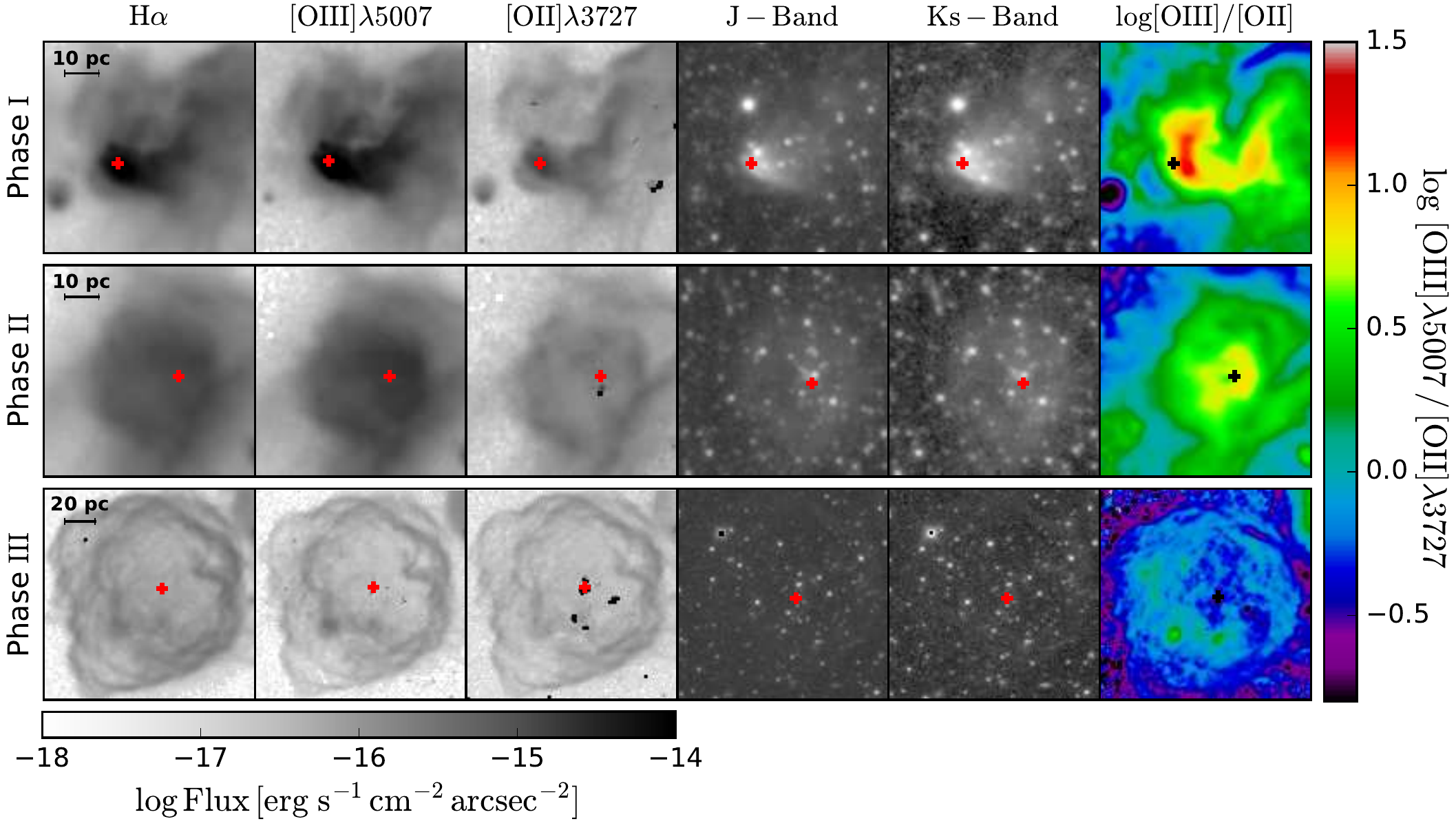}
		\caption{{\em Left}: A schematic of our emission model. During Phases I and II, emission is produced by both the shell and the undisturbed natal cloud. During Phase III, however, the cloud has been completely swept up and so all of the emission comes from the shell. Note that the gas density in the central bubble is very low, so we do not expect this region to contribute significant emission in the lines of interest in this study. 
		{\em Right}: Observations of star-forming regions in various evolutionary phases, traced by \Ha, \OIII and \OII nebular emission. Near-infrared emission reveals a similar cloud morphology to the line emission, demonstrating that the morphology that we can infer from the line emission is not strongly affected by attenuation. Observations are described in \citet{Rousseau-Nepton2018}.} 
		\label{fig:Cartoon}
	\end{center}
\end{figure*} 

The {\sc warpfield-emp} framework consists of three components: {\sc warpfield} \citep{Rahner2017a,Rahner2018}, used to model the evolution and physical structure of the H{\sc ii} region and PDR; {\sc cloudy} \citep{Ferland2017}, used to determine the chemical state of the gas, the dust temperature, and the emission properties of the gas and dust; and {\sc polaris} \citep{Reissl2016}, used to account for the effects of reddening on the emergent radiation. As all three codes are described in detail elsewhere, here we will only give a brief outline of each, with a focus on any modifications we have made compared to the previously published versions. We will also discuss how we couple the three codes together in the {\sc warpfield-emp} framework.

\subsection{{\sc warpfield}}
\label{sec:wf}
{\sc warpfield} solves the equations of motion for a 1D spherical shell\footnote{Although our 1D approach is primarily motivated by the desire to avoid the computational costs of a full 3D approach, which are too high to allow one to properly survey the relevant parameter space, there is also a physical justification underpinning this. The internal structure of \hii regions traced by line ratios reveals them to be largely regular, mostly circular objects when studied in ionization parameter mapping line ratios \citep{Pellegrini2012} in the Magellanic clouds or NGC~628 \citep{Rousseau-Nepton2019}. Thus despite the variability in initial conditions, the cumulative effects of all feedback seem to wash out inhomogeneities and often produces spherical objects that should be described well by our 1D models.} evolving under the influence of stellar winds, supernova energy injection, radiation pressure and gravity. It assumes that the evolution of the region affected by feedback can be separated into several distinct phases, illustrated schematically in Figure~\ref{fig:Cartoon}. 

In Phase I, the shell is filled with hot gas from shocked stellar winds, and its expansion is driven by the pressure difference between the hot interior gas and the much colder ambient medium. During this phase, the effects of gravity and radiation pressure are negligible, leaving one a relatively simple differential equation to solve for the shell motion \citep{Weaver1977,Bisnovatyi-Kogan1995,Rahner2017a}. Phase I comes to an end once the bubble loses its hot gas, either because the gas cools, which occurs after a cooling time $t_{\rm cool}$, during shell-fragmentation \citep{Rahner2019}, or because the bubble ``bursts'', allowing the hot gas to escape through lower density channels \citep{Rogers2013}. In our 1D treatment, we cannot solve explicitly for effects like the bursting of the bubble, which are influenced by the 3D structure of the cloud, and so by default we assume that the bubble bursts once the shell has swept up the entirety of the initial cloud, i.e.\ once the shell radius is equal to the initial cloud radius. This occurs at a time $t_{\rm sweep}$, which we can readily determine from our 1D treatment. Phase I therefore comes to an end after a time $t = {\rm min}(t_{\rm cool}, t_{\rm sweep})$.\footnote{Users of {\sc warpfield} are free to configure different burst parameters relative to the natal cloud size.}

Following Phase I, the shell either proceeds to Phase II (if $t < t_{\rm sweep}$) or directly to Phase III (if $t \geq t_{\rm sweep}$). In Phase II, the shell continues to sweep up material from the surrounding molecular cloud. The expansion of the shell is driven by the ram pressure exerted on the shell by stellar winds and supernovae, and by the radiation pressure acting on the shell, but at this point we also start to account for the counteracting effect of gravity, both due to the gravitational attraction of the central cluster on the shell and also the shell's own self-gravity. Phase III is very similar to Phase II in terms of the force balance acting on the shell, with the key difference being that during Phase III, the original cloud has been entirely swept up, with the shell now assumed to be expanding into the low density warm neutral medium (WNM). 

There are two possible fates for the shell. If gravity begins to dominate, which often happens as the central stellar cluster ages and feedback becomes less effective, then the shell will begin to recollapse. In the original {\sc warpfield} implementation \citep{Rahner2017a}, we simply terminated the calculation when the shell had recollapsed to a radius of 1~pc. In {\sc warpfield-emp}, however, we adopt instead the approach used in \citet{Rahner2018} and assume that recollapse triggers a new burst of star formation, which for simplicity is assumed to have the same star formation efficiency as the original burst. 
The other possible fate is continued expansion without recollapse, which will occur once the kinetic energy of the shell exceeds its gravitational potential energy. In this case, we follow the expansion of the shell until its maximum density falls below $1 \, {\rm cm^{-3}}$ for an extended period of time (more than 1~Myr). Following this, we consider the shell to have dissolved. In the analysis in Section~\ref{sec:results1}. we also ignore models where the shell has expanded to more than 200\,pc since at this point the influence of environmental effects such as galactic shear or interactions with other clouds would start to become important -- issues which are currently ignored in \textsc{warpfield}.

As part of {\sc warpfield}, we solve not only for the dynamics of the shell, but also for its structure, and the density gradients driven by radiation pressure, as well as the natal cloud structure. This strongly influences the effectiveness of radiation pressure driving. To do this, we assume that the ionized and neutral/molecular portions of the shell are in quasi-hydrostatic equilibrium, with the equation of state given by \citet{Abel_2005} and  \citet{Pellegrini2007}. By default, {\sc warpfield} accounts only for thermal pressure support when solving for the shell structure. However, the code offers the option of also including the effects of magnetic and/or turbulent pressure support, as discussed at greater length in Appendix~\ref{appendix:Bfield}. In the results presented in this paper, we do not make use of these options in order to minimize the number of free parameters.

The four main input parameters to any {\sc warpfield} model are: the mass of the gas cloud ($M_{\rm cl}$); the initial mass of the stellar cluster, ($M_{*}$), which can either be specified directly or in terms of a star formation efficiency $\epsilon$, in which case $M_{*} = \epsilon M_{\rm cl}$; the density of the cloud ($n_{\rm cl}$); and the metallicity. In the models presented later in this paper, we keep the metallicity fixed at the solar value, but explore the effects of varying $M_{\rm cl}$, $\epsilon$ and $n_{\rm cl}$. We assume that the stars within the star cluster follow the Kroupa initial mass function \citep{Kroupa2002} where the number of stars $N$ with initial mass $M$ is given by
\begin{align} \label{KroupaIMF}
  \frac{\diff N}{\diff \log (M)} \propto \begin{cases}
      M^{-0.3}, & 0.08\,\rm{M}_{\odot} \leq M \leq 0.5\,\rm{M}_{\odot}\\
      M^{-1.3}, & 0.5\,\rm{M}_{\odot} < M \leq 120\,\rm{M}_{\odot} .
    \end{cases}
\end{align}

\subsection{{\sc Cloudy}}
{\sc Cloudy} is a non-local thermodynamic equilibrium (NLTE) spectral synthesis and plasma simulation code developed over the past three decades by G.~Ferland and collaborators. In {\sc warpfield-emp}, we use the latest major version of {\sc Cloudy}, version C17, described in detail in \citet{Ferland2017}. We use {\sc Cloudy} to determine the emissivities of a large set of emission lines as a function of position within the shell and the surrounding undisturbed cloud. {\sc Cloudy} also provides us with a description of how the corresponding opacities vary within the shell and the cloud. This data is then passed on to {\sc polaris} to produce the final emission line maps, as described in Section~\ref{sec:polaris} below.

For shells in Phase III, the procedure is straightforward. In this case, the shell has swept up the entirety of the surrounding cloud and so we need only consider emission from the shell itself. To do this, we pass the following information to {\sc Cloudy}: the frequency-dependent flux from the stellar cluster that is incident on the inner edge of the shell, the radial distance from the central cluster to the shell, the total pressure (ram plus thermal) acting on the inner shell boundary, the velocity of the shell, the total shell mass, and the metallicity. With the exception of the frequency-dependent flux, which we generate using the time-dependent cluster model discussed in Section~\ref{sec:starcluster} below, all of these parameters either come directly from our {\sc warpfield} model for the shell or from the initial conditions for the calculation. 

Given this information, {\sc Cloudy} then solves for the structure of the shell by breaking it up into a series of thin concentric shells (``zones'' in the terminology used in the {\sc Cloudy} documentation). The thickness of each of these zones is determined internally by {\sc Cloudy} and is set by the requirement that the physical conditions across each zone must be close to constant. Consequently, the zone thickness can vary significantly as a function of distance from the cluster -- for example, the zones in which the ionization front are located must be very thin in order to properly capture the transition from ionized to neutral gas. The density in each zone is determined by {\sc Cloudy}, using the same assumption of quasi-hydrostatic equilibrium as in {\sc warpfield}.\footnote{The difference between the simple cooling treatment used in {\sc warpfield} and the more sophisticated treatment in {\sc cloudy} can lead to minor differences in the shell structure. This issue is discussed in more detail in Appendix~\ref{appendix:diffs}.} {\sc cloudy} also solves self-consistently for the thermal structure of the gas, its chemical composition, and the emissivities and opacities for a large set of different emission lines. 

For shells in Phase I or Phase II, which have not yet swept up all of the natal cloud, there is one additional step to the procedure. Once we have determined the shell structure and emission, we then carry out a second {\sc Cloudy} calculation to determine the emission from the natal cloud. In this case, the incident flux is taken from the output of the {\sc Cloudy} calculation for the shell.\footnote{We do not currently account for the effects of radiation from outside of the cloud, in the form of the interstellar radiation field, since this will usually be unimportant compared to emission from the central stellar cluster.} The cloud is assumed to have zero bulk velocity relative to the cluster, and to have a turbulent velocity dispersion set by the requirement that the cloud be initially in virial equilibrium. The other parameters that are required -- the mass, density and size of the cloud, the location of the outer edge of the shell, and the metallicity -- come directly from the {\sc warpfield} calculation or the initial conditions. As in the case of the shell calculation, {\sc Cloudy} breaks the cloud up into a series of zones and computes the emissivities and opacities for each zone.

\subsection{{\sc Polaris}}
\label{sec:polaris}
\begin{figure}
	\includegraphics[width=0.5\textwidth]{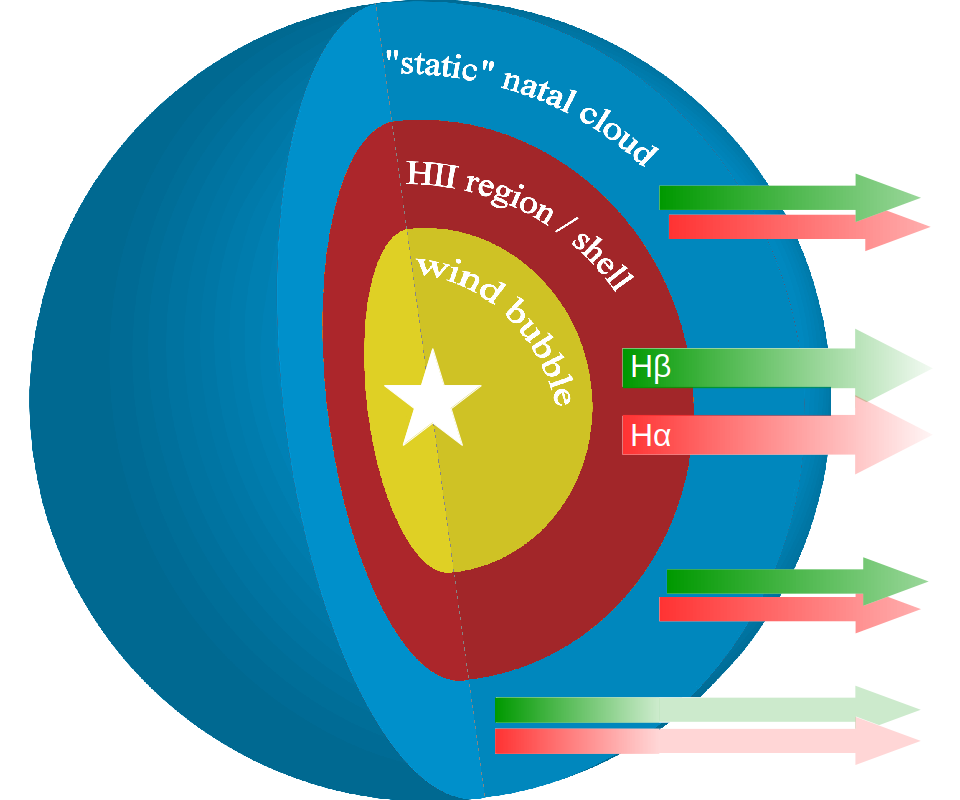} 
	\caption{A schematic of an embedded \hii region, surrounded by a partially ionized cloud. The net emission in \Hb, more blue and attenuated by dust than \Ha, may be dominated by the static component, while \Ha emission from a \hii region contributes measurably to the net flux from the cloud. In such a scenario, common to many of our models, the Balmer decrement is affected by differential attenuation and therefore, it has no simple physical meaning and any reddening correction derived from it may be highly inaccurate.}
	\label{fig:SaturatedAv}
\end{figure}

The time-dependent evolution of density and column density is large in our models. The resulting internal extinction by dust plays a major role in determining the emergent lines intensity and therefore the ratios. For example, the ratio of the ionizing photon production rate, $Q_0$, to the emergent \Ha flux varies over 9 orders of magnitude between ultra-compact \hii regions with $A_{\rm V} \geq 20$ and unembedded \hii regions with $A_{\rm V} \ll 1$. Our model predicts regimes where the illuminated portion of the shell that is the  \hii region is heavily obscured by surrounding material, meaning that it is possibly unobservable in many tracers. Further complicating the picture, radiation making it out of the \hii region into the lower density ambient cloud can produce low ionization emission that can dominate over the emergent light from the \hii region (see Figure~\ref{fig:SaturatedAv}). It is therefore extremely important to account for the impact of dust extinction. To do this, we make use of a modified version of the {\sc polaris} RT code\footnote{\url{http://www1.astrophysik.uni-kiel.de/~polaris/}} \citep{Reissl2016}. 

{\sc polaris} was originally developed to model polarized dust continuum emission from 3D dust distributions using the Monte Carlo method \citep[see e.g.][for some example applications]{Reissl2017,Reissl2018}, but more relevant for our purposes is its ray-tracing mode. It has been used for various applications, such as state-of-the-art treatments of dust grain alignment physics \citep[][]{Brauer2016,Reissl2016,Reissl2017,Seifried2018,Reissl2018}, atomic and molecular line RT including the Zeeman effect \citep[][]{Brauer2017,Reissl2018}, as well as synchrotron polarization and Faraday rotation \citep[][]{Reissl2018prep} on variable 3D grid geometries (adaptive octree, spherical, cylindrical, and native Voronoi).

In ray-tracing mode, {\sc polaris} solves the multi-frequency equation of RT on a set of parallel rays passing through a 3D grid, self-consistently accounting for the absorption and emission of radiation by dust. When used for line transfer, it can either compute the line emissivities internally, using e.g.\ the local thermodynamic equilibrium (LTE) or large velocity gradient (LVG) approximations to determine the local levels population, or it can adopt pre-computed values per grid cell supplied by the user. In {\sc warpfield-emp}, we use the latter option, supplying the emissivities computed by {\sc cloudy} as an input to {\sc polaris}. In addition, we also supply {\sc polaris} with the line and continuum opacities determined by the {\sc cloudy} calculation\footnote{In practice, for the diagnostics examined in this paper, the line opacities are very small and the dominant contribution comes from the dust. However, our method does not require this to be the case and can also be applied to produce maps of tracers with high line opacities, such as CO.}. 

{\sc polaris} supports several different types of 3D grid, but in {\sc warpfield-emp} the obvious choice is a 3D spherical grid, as in this case the transfer of information from the 1D model to the 3D grid is trivial. When setting up our spherical grid, we ensure that the length of each radial bin $dr$ matches the thickness of the corresponding {\sc Cloudy} zone, allowing us to map data directly from the zones to the radial bins. This required us to modify {\sc polaris} to support spherical grids with arbitrarily-sized radial bins and potentially with large changes in thickness between bins. This is important for resolving the narrow emission line region near the ionization front, without requiring us to use a prohibitively large number of radial bins. We then carry out a ray trace along a set of parallel rays passing through the grid, allowing us to project the 3D model onto a 2D detector. When projecting from 3D onto our 2D detector, we use a uniform pixel matrix. Here, we make use of {\sc polaris}'s recursive sub-pixelling technique, which splits up rays as necessary to ensure that every cell in the 3D grid is sampled by at least one ray, thereby ensuring that we do not lose information from regions smaller than the size of the pixels. Finally, for carrying out the ray-tracing, cubic spline interpolation is used to interpolate values from the grid onto the rays, and the Runge-Kutte-Felhberg (RKF45) solver with an inbuilt error and step size correction is used to solve the RT equation along the rays. Since the radial distances $dr$ between adjacent grid cells are already optimized by {\sc cloudy} we limit the RKF45 solver to refine its step size only down to $0.1\times dr$ without losing accuracy. This enables us to ray-trace the enormous number of {\sc cloudy} models within a reasonable time with an error below $1\ \%$.

With this approach, we can produce maps of velocity-integrated surface brightness of any line or continuum tracers, provided that they are included in the {\sc cloudy} model. Alternatively, we can produce full position-position-velocity (PPV) cubes of the  emission line. In the latter case, the line profiles are calculated using a specified grid of velocity channels and accounting for the line-of-sight projected velocity, turbulence and thermal broadening of each line. In calculations that include both the shell and the remaining natal cloud, we make the simplifying assumption that there is no line overlap between the two components when the shell expansion velocity is much greater than the turbulent velocity dispersion of the cloud ($v_{\rm exp} \gg \sigma_{\rm turb}$).

Optionally, the ray tracing can be perform with no dust opacity, recovering what would be the line emission without internal extinction. 

\begin{figure}
	\centering
	\includegraphics[width=1.0\linewidth]{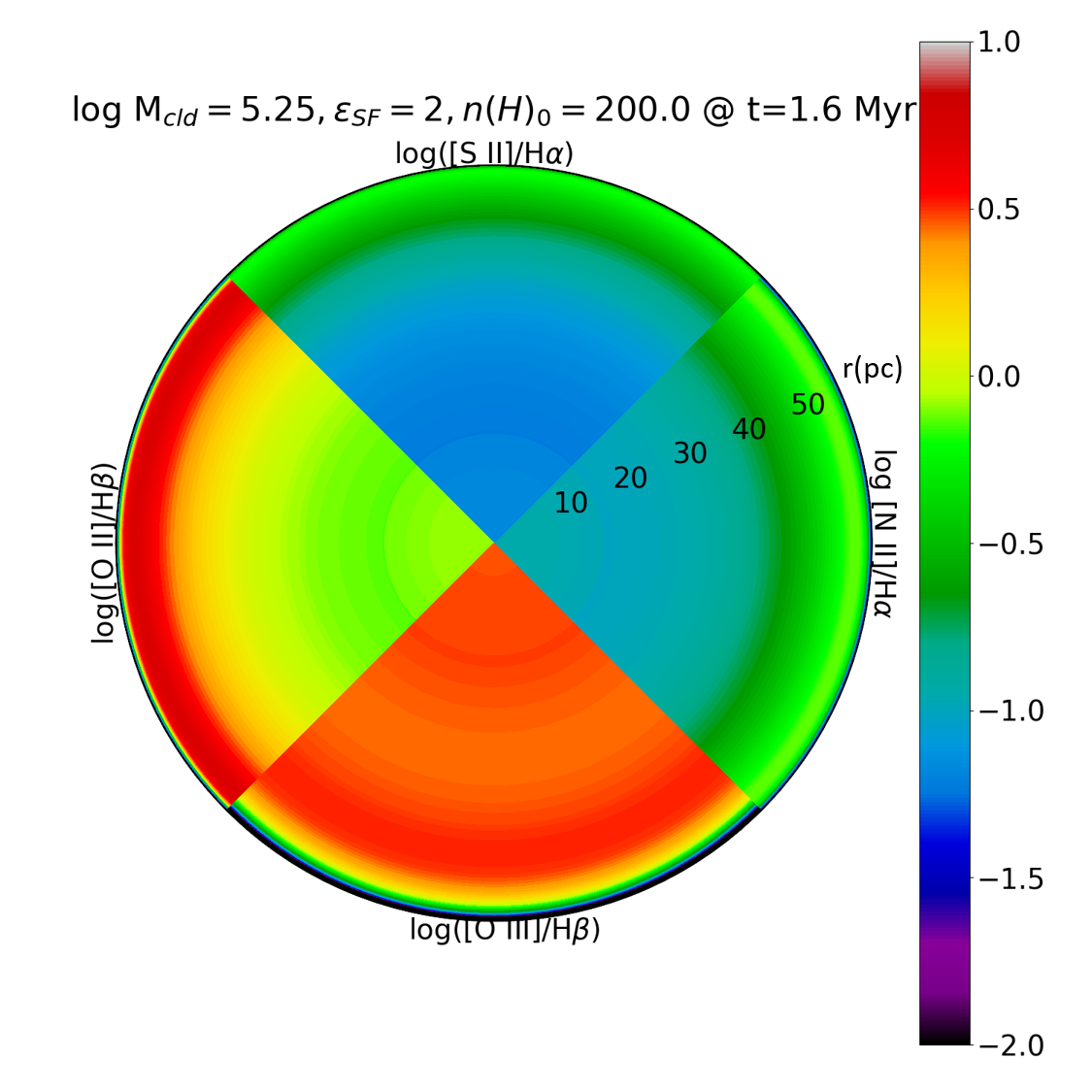}
	\caption{2D projected emission line ratio ratio maps of an example model with cloud mass $10^{5.25} \, {\rm M}_\odot$, initial cloud density $n_{\rm H}= 200 \, {\rm cm}^{-3}$, and star formation efficiency $\epsilon = 2$\%. The calculation of emergent fluxes includes internal reddening. The values are shown at a time $t = 1.6$~Myr after the formation of the central stellar cluster. The different panels show the ratio of the \OIII 5007\AA\; line to \Hb (bottom); the ratio of the \SII 6716+6731\AA\; doublet to \Ha (top); the ratio of \NII to \Ha (right) and the ratio of the \OII doublet to \Hb  (left). The image is projected in 2D at a radial resolution equal to the native {\sc cloudy} grid, projected along the z-axis.
	}
	\label{fig:combination-2d-projection-m6}
\end{figure}

In Figure~\ref{fig:combination-2d-projection-m6}, we show an example of the kind of output that we can produce with {\sc warpfield-emp}. Each quadrant shows a section of the map we obtain by taking the ratio of the surface brightness of some observational tracers.
The ratio \OIII/\Hb, shown in the lower quadrant, traces regions with a large flux of energetic ionizing photons and hence is high throughout much of the projected bubble. However, it drops off dramatically toward the edge of the bubble as the ionizing flux drops, and the internal density increases. The ratio \SII/\Ha, shown in the top quadrant, has very different behavior. It is low within most of the low density bubble (as most of the sulphur present has been ionized to S$^{2+}$), but increases dramatically as one nears the dense shell. This behavior is characteristic of optically thick \hii regions, and results in a pronounced increase in the \SII/\OIII ratio at the \hii region boundary that is observed in many resolved \hii regions \citep[see e.g.][]{Pellegrini2012}. The layer with enhanced \SII/\Ha is thicker than the shell itself, which is only $\sim 0.05$~pc wide at this point in the evolution of the bubble. We also see similar morphology in the predicted \NII/\Ha flux (right) and \OII/\Hb flux (left). The region with elevated \NII/\Ha is thicker than that with elevated \OII/\Hb, due to the higher ionization potential, even though the total intensity of \OII is higher owing to the higher abundance of oxygen relative to nitrogen. It is important to note that much of the details of radiation transfer, such as the inner boundary, are difficult to observe directly.

\subsection{Star Clusters}
\label{sec:starcluster}

\subsubsection{Individual Cluster Evolution}
In order to allow us to focus in this paper on the effect of a time-dependent, dynamic cloud structure on predictions for emission line diagnostics, we adopt a very simple star cluster model for the simulations presented in this paper. We assume that a given burst of star formation forms a star cluster with mass $\epsilon M_{\rm cl}$, where $\epsilon$ is the star formation efficiency and $M_{\rm cl}$ is the mass of gas in the cloud. The star formation efficiency and the initial mass of the cloud are input parameters of our model. For bursts of star formation occurring following re-collapse of the shell, we adopt the same star formation efficiency as for the initial burst, and compute the mass of gas in the cloud by subtracting the mass of stars already formed from the initial cloud mass. The stellar population of the cluster formed by a given burst of star formation is then simply scaled from a fully sampled $M = 10^6 M_{\odot}$ cluster. As in our previous {\sc warpfield} papers, we use the time-dependent evolution of a cluster calculated from {\sc Starburst99}\footnote{\url{http://www.stsci.edu/science/starburst99/docs/default.htm}} \citep{Leitherer1999,Leitherer2014}, using the Geneva stellar evolution tracks for rotating stars \citep{Ekstrom2012}. In a subsequent paper, we will introduce the complexity of stochastically sampled clusters when we consider populations of star-forming regions, but for now we focus only on predictions and degeneracies introduced with a fixed cluster template.

\subsubsection{On-going Star Formation}
While an individual stellar cluster may be considered ``simple'', the ionizing spectral energy distribution (SED) of our models is not. Unlike typical photoionization grids employing a single assumed cluster SED, our SED not only evolves in time with our physical conditions, but the inclusion of subsequent bursts of star formation following re-collapse of the shell leads to complex stellar populations, composed of multiple massive clusters with intermediate ages (few Myr), dominated at times by O- and/or B-stars. Normally the clusters with O-stars dominate the ionizing energy regime, but that is not guaranteed. Multiple older clusters with ages ranging from 2-10~Myr can be the dominant source of radiation in other frequency bins. This is especially true below the hydrogen ionization limit, and can greatly affect the ratios of low-to-mid ionization potential lines to high-ionization potential emission. 

\begin{figure*}
	\begin{center} 
		\includegraphics[width=0.85\textwidth]{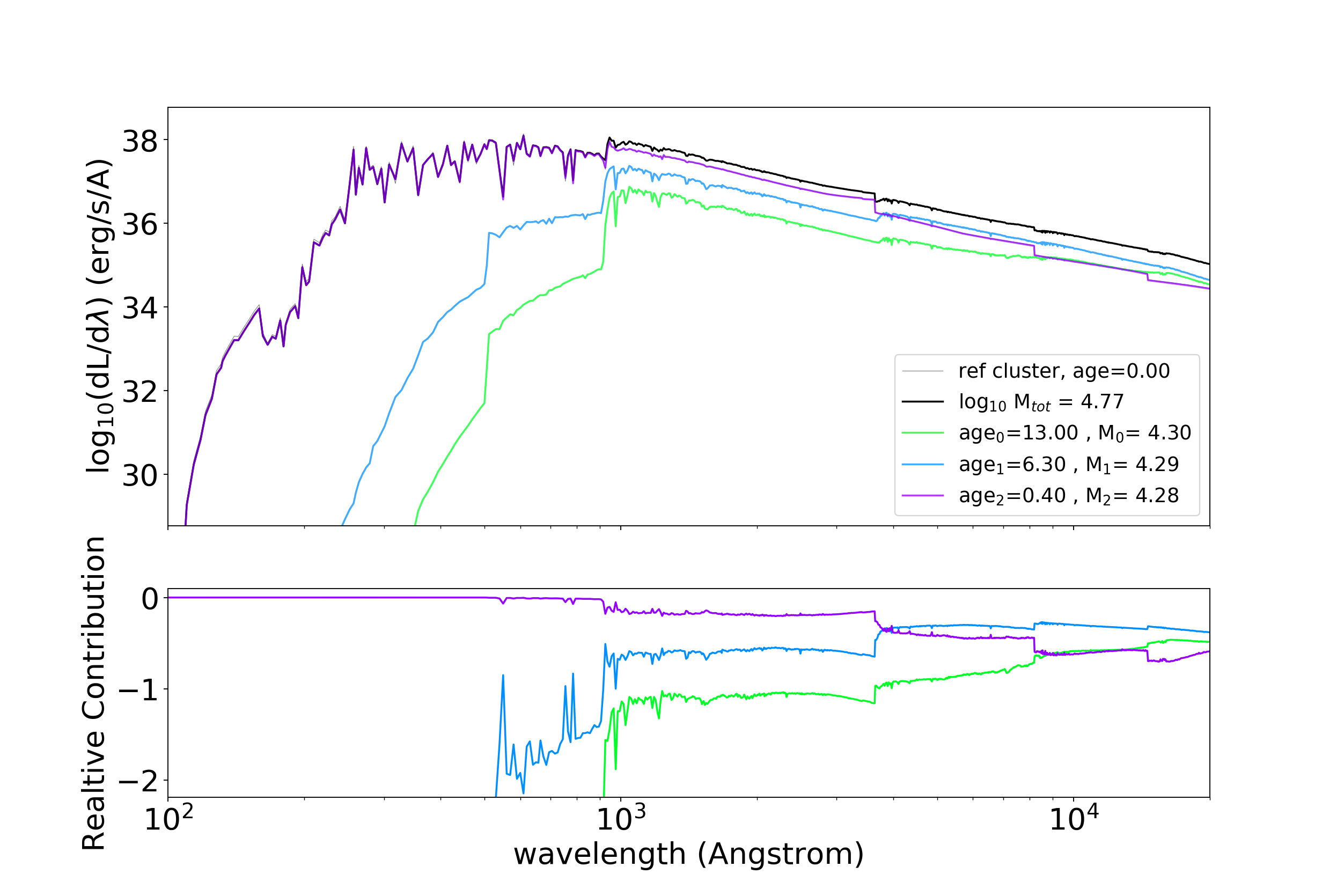} 
		\caption{Example of a cluster with a complex star formation history. In this model, which has a cloud mass $M_{\rm cl} =  10^6 \, M_\odot$, a cloud density $n_{\rm H} = 500 \, {\rm cm^{-3}}$ and a star formation efficiency $\epsilon = 0.02$, the cluster at the time depicted has undergone three bursts of star formation -- an initial burst, and two subsequent bursts associated with shell recollapse.  In the top panel, we show the spectral energy distributions (SEDs) of the three distinct stellar populations, at a time when their ages are $t = 0.4, 6.3$, and 13~Myr, respectively. The SED of the oldest cluster at time $t = 0$ is shown with the gray line. The total mass of the cluster (in log units) and the masses of the individual stellar populations are listed in the inset. We see that the ionizing luminosity and the B-band are dominated by emission from the youngest cluster, while the light at red wavelengths and in the IR is dominated by the contribution of the older clusters. This is also highlighted in the bottom panel, where we show the fractional contribution to the total luminosity at each wavelength made by each separate stellar population.}
		\label{fig:SpectralEvo}
	\end{center}
\end{figure*}

As an example, we show in Figure~\ref{fig:SpectralEvo} the case of a cluster that has undergone three separate bursts of star formation: an initial burst, plus two subsequent bursts associated with shell re-collapse. At short wavelengths -- particularly shortwards of 912~\AA\, -- the SED of the cluster is dominated by the contribution from the youngest stellar population, while at long wavelengths, the two older stellar populations dominate. 

\section{Time-dependent Diagnostic Diagrams} \label{sec:results1}
One of the major strengths of the {\sc warpfield-emp} framework is the ability it gives us to examine the time-dependent behavior of the emission from an evolving \hii region and PDR in a physically self-consistent fashion. As a star cluster ages, its SED changes \citep{Leitherer1999}, and so the relative strengths of the emission lines associated with the ionized gas also change. However, the strengths of these lines are also sensitive to the density of the emitting gas, which changes as the \hii region evolves, as well as other variables such as the distance of the ionized gas from the cluster. Moreover, these variables are correlated, since the shell density, radius etc.\ depend to a great extent on the time history of feedback from the central cluster. Therefore, approaches in which these variables are varied independently are likely to yield misleading results.

As an example of the power of the {\sc warpfield-emp} approach, we examine the example of a pair of widely-used observational diagnostics of ionized gas, the ratios of the \OIII 5007\AA~line to the \Hb line (\OIII/\Hb) and the ratio of the \NII 6583\AA~line to the \Ha line (\NII/\Ha). The use of these ratios, rather than the strengths of the \OIII and \NII lines themselves, has two big advantages. Firstly, it allows one to meaningfully compare regions with very different ionizing photon luminosities, as changes in $Q_{0}$ will to zeroth order change \OIII and \Hb or \NII and \Ha by the same amount. Secondly, since \OIII lies close to \Hb in terms of wavelength, dust extinction has a similar effect on the strength of both lines. The same is true for \NII and \Ha, for the same reason. Therefore, the values of the \OIII/\Hb and \NII/\Ha line ratios produced by a given parcel of gas are largely unaffected by dust extinction. 

\begin{figure*}
	\centering
	\includegraphics[width=0.95\linewidth]{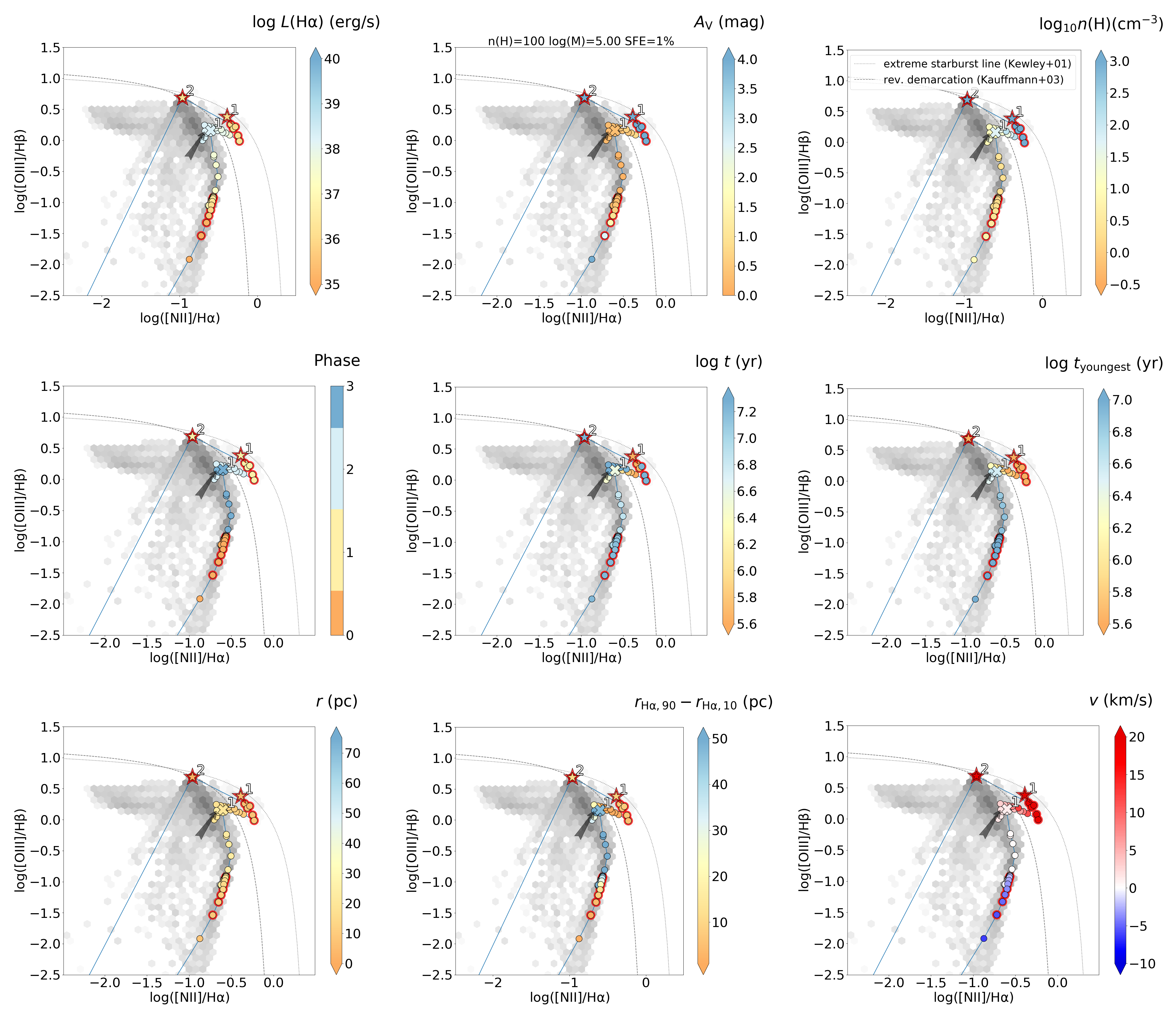}
	\caption{
	Temporal evolution of the  \OIII/\Hb and \NII/\Ha line ratios in an example model ($M_{\rm cloud} = 10^{5.0} M_\odot$, $n_{\rm H} = 100 \, {\rm cm}^{-3}$, SFE = 1\%). The line shows the evolutionary track, while the circles represent the predictions at each non-uniform time step. In each panel color represents an observable, a cloud property, or property of the system. For each star formation event the first data point after the star formation event (at 0.1\,Myr) and the occurrence of first supernova (at 3.6\,Myr after cluster formation) are marked by a star symbol and cross, respectively, with a number indicating which sub-cluster the event belongs to. The end-point of the model is indicated by an arrow. Symbols outlined in red indicate times at which the system is unlikely to be observable, given typical sensitivity limits for extragalactic observations.
	For reference, we also indicate the criteria proposed by \citet{Kewley2001} (dotted line) and \citet{Kauffmann2003} (dashed line) for distinguishing between star-forming regions and AGN. The results from our full set of models at every output time are shown in the gray-scaled hexbin histogram in the background. Contributions to this histogram are weighted by the size of the time-step.
	From left to right and top to bottom, the panels show: 
	Emergent, dust-attenuated \Ha luminosity; visual extinction ($A_{\rm V}$) along the line of sight, computed assuming $R_{\rm V} = 3.1$, total H nuclei number density at the wind-bubble--\hii region interface; {\sc warpfield} evolutionary phase, with $0$ corresponding to re-collapse; age of system from the beginning of the simulation; age of the youngest star cluster (accounting for re-collapse);
	radius of the bubble/shell boundary in pc; difference of the radii (as measured from the bubble/shell boundary) containing 90\% and 10\% of the integrated \Ha luminosity ($r_{\Hamath, 90}$, $r_{\Hamath, 10})$; and lastly the shell velocity, with positive values corresponding to expansion and negative to re-collapse.
	}
	\label{fig:BPT_time_evol}
	\end{figure*}

One of the most common uses of these line ratios is to identify the mechanism responsible for ionizing the gas. \citet{Baldwin1981} demonstrated that if one constructs a plot of \OIII/\Hb vs.\ \NII/\Ha for a given galaxy -- an example of what has become known as a BPT diagram -- then regions where photoionization by stars dominates are found in a clearly distinct part of the plot from regions dominated by a hard ionizing spectrum (e.g.\ an AGN) or by shocks. More recently, \citet{Kewley2001} showed that there are clear limits on how large \OIII/\Hb and \NII/\Ha can be in a region dominated by stellar photoionization, even in the most extreme starburst. These are indicated as the dotted line in each panel of Figure~\ref{fig:BPT_time_evol}. In practice, most real star-forming galaxies sit somewhat below and to the left of this line in the BPT diagram, in the region bounded by the dashed line in each panel of Figure~\ref{fig:BPT_time_evol} \citep{Kauffmann2003}.

To explore how we expect the \OIII/\Hb and \NII/\Ha ratios to change as a function of time in evolving \hii regions, we have carried out an extensive series of simulations using {\sc warpfield-emp}. As summarized in Table~\ref{tab:model_grid}, we have examined models with three different initial densities, six different star formation efficiencies, and cloud masses ranging from $M_{\rm cl} = 10^{5} \: {\rm M_{\odot}}$ to $10^{7.25} \: {\rm M_{\odot}}$ in 0.25~dex increments, yielding a total of 180 different models. For simplicity, in this initial exploration of parameter space, we do not vary the metallicity, keeping it fixed at ${\rm Z} = {\rm Z_{\odot}}$ in every run. This helps to limit the number of models we need to run, and also allows us to explore how much variation we expect in the \OIII/\Hb and \NII/\Ha ratios at fixed metallicity.

\begin{table}
	\caption{Range of parameters examined in the models discussed here} 
	\label{tab:model_grid}
	\small 
	\centering 
	\begin{tabular}{lcl} 
		\toprule[\heavyrulewidth]\toprule[\heavyrulewidth]
		Parameter & Unit & Values \\ 
		\midrule
		 {$n_{\rm H}$} &  [${\rm cm}^{-3}$] & 100, 200, 300, 500 \\
 		 {$\log_{10} M_{\rm cl}$} & [{${\rm M}_{\odot}$}] & 5.0, 5.25, ..., 7.00, 7.25 \\
                  SFE & [\%] & 1, 2, 4, 6, 8, 10 \\
		\bottomrule[\heavyrulewidth] 
	\end{tabular}
\end{table}

To limit the model output data volume and avoid unnecessary computational work, we do not re-run {\sc cloudy} and {\sc polaris} after every {\sc warpfield} timestep. Instead, we delay re-running {\sc cloudy} until the physical conditions in the cloud or the properties of the star cluster have changed sufficiently to result in an appreciable change in the emission. In practice, this means that we re-run {\sc cloudy} whenever the shell density at the inner shell boundary, shell radius, shell mass or the ionizing photon flux change by the fractional amounts shown in Table~\ref{tab:model_delta_limits}. We also recalculate models when the {\sc warpfield} model changes evolutionary phases, or if more than 0.5~Myr has elapsed since the last calculation.

\begin{table}
	\caption{Maximum parameter variation between adjacent {\sc cloudy} models} 
	\label{tab:model_delta_limits}
	\small 
	\centering 
	\begin{tabular}{lcccc} 
		\toprule[\heavyrulewidth]\toprule[\heavyrulewidth]
		Parameter & {$n_{\rm H, shell}$} & {$r_{\rm shell}$} & $M_{\rm shell}$ & {$Q_0$} \\ 
		\midrule
		$\Delta_{\rm max}$(\%) & 20 & 10 & 20 & 20 \\
		\bottomrule[\heavyrulewidth] 
	\end{tabular}
\end{table}

In Figure~\ref{fig:BPT_time_evol}, we show the results from an example model with $M_{\rm cl} = 10^{5.00} {\rm M}_\odot$, $n_{\rm H} = 100 \, {\rm cm}^{-3}$ and SFE = 1\%. In grey scale we show the full distribution of our models. 
Further examples are shown in Appendix~\ref{appendix:SupPlots} for different cloud properties. Both observables, such as the \Ha luminosity, and intensive properties, such as the density at the bubble/\hii region boundary, evolve rapidly with time. At early times the cluster luminosity is relatively constant, and this evolution is driven by the change in internal extinction. The density plotted is the density at the bubble boundary, and is set by the internal hot wind pressure, or after the bubble has burst, cooled, or fragmented, the direct momentum of the free streaming wind. For the ionization front density, as with all radiation-pressurized nebulae, there is a relationship between density diagnostics and the photon flux incident on the cloud. This produces a fundamental relationship between star-forming events and density, but one that is less sensitive to initial conditions and more a reflection of the current nebular structure and cluster age. Thus, different trends with ionized gas densities identified in nearby galaxies in \citet{Herrera2016} should be studied in the context of full population synthesis, and are beyond the scope of this immediate paper. It is sufficient to say that these trends are related to, but not caused by environment. 

In the second row, we plot the {\sc warpfield} evolutionary phase and ages of the models. 
We see immediately that both line ratios change substantially as the cluster and the shell evolve, particularly the \OIII/\Hb ratio. To help understand this we introduce a metric to quantify the relationship between geometry and emission, $r_{\rm obs, fraction}$, to represent the radius from the bubble boundary where the cumulative luminosity of a particular observable reaches a certain fraction of the total the model produces.

Two uses of the ratio are shown in the last two panels of Figure~\ref{fig:BPT_time_evol}, where we show the values of $r_{\Hamath, 90} - r_{\Hamath, 10}$ which is a proxy for the observable thickness of the \hii region, as traced by \Ha, a better way of quantifying whether the shell is ``thick'' (in which case the value of the thickness is large compared to the inner radius of the bubble) or ``thin'' (in which case the thickness is small compared to the size of the bubble). We can also use \OIII and \Ha because their strong differences in ionization potential guarantees that we are tracing the extremes of the \hii ionization structure. 

At early times, during the first 2\,Myr, the shell expands at high velocities, $v > 10$\,km\,s$^{-1}$, gradually slowing down as it sweeps up more and more material. In the particular example shown, the fast growth of the shell coincides with an increase in \OIII/\Hb at very early times. The increase comes as the adiabatic expansion decreases the internal pressure, leading to a decrease in the density of the ionized gas in the shell. This density is initially very high because the high internal pressure in the bubble strongly compresses the surrounding ionized shell. As the shell accelerates, the internal volume increases, leading to a decrease in the density of the shell and an increase in the \OIII/\Hb ratio. 

At later times, the emission rate of ionizing photons starts to decrease when the most massive stars die, and the \OIII/\Hb ratio drops again. The shell expansion velocity, which has decreased to $\sim 5$~km\,s$^{-1}$, increases again as the ram-pressure of the first SN explosions drives the gas away from the star cluster. Shortly afterwards, the nebular emission reaches its maximum \Ha luminosity of $\sim 10^{38}$\,erg\,s$^{-1}$. After approximately 10\,Myr, as the shell extends over an ever increasing  volume, the density in the shell drops so low that it becomes easier to produce \OIII again even though $Q_0$ continues to drop. However, we note that the evolution of the presented example model should not be taken to be representative of all star-forming regions. Objects with a stronger starburst, where the cloud is quickly destroyed by feedback, and objects with a lower SFE where consecutive star formation events will occur, can -- and often will -- have a very different evolution in a BPT diagram, as the additional examples shown in Appendix~\ref{appendix:SupPlots} make clear.

A final note on recollapsing objects: We often find that the density of the shell in a recollapsing bubble decreases even as the radius becomes smaller. This is a reflection that the feedback is decreasing faster than the gas can compress, manifesting as a thicker \hii region with higher filling factor and lower average density. In some time series of more massive clouds, this balance can lead to an increase in \OIII/\Hb.

Although informative, plots such as those in Figure~\ref{fig:BPT_time_evol} that show where the emission from a single star-forming cloud falls in the BPT diagram throughout the lifetime of the cloud are also potentially misleading, since not all evolutionary stages are observable. Early on, before the whole of the natal cloud is swept up into the shell, the foreground dust extinction due to the undisturbed portion of the cloud can be large. This has little impact on the location of the points in the plot, since the values of the \OIII/\Hb and \NII/\Ha line ratios are insensitive to extinction. However, the absolute brightness of the lines is strongly affected and sufficient dust may render them undetectable given any plausible amount of observing time. At late times, on the other hand, the \hii region becomes faint not because of dust extinction but simply because by this time, most of the massive stars powering the emission have already died. In addition, the emission that is present is spread over an increasingly large area of sky, owing to the expansion of the shell, further reducing its surface brightness. In order to compare results from {\sc warpfield-emp} with observations of real \hii regions in a meaningful fashion, it is necessary to account for both of these effects. We discuss how we do this in Section~\ref{sec:obscomp} below.

\begin{figure*}
    \centering
    \includegraphics[width=0.95\textwidth]{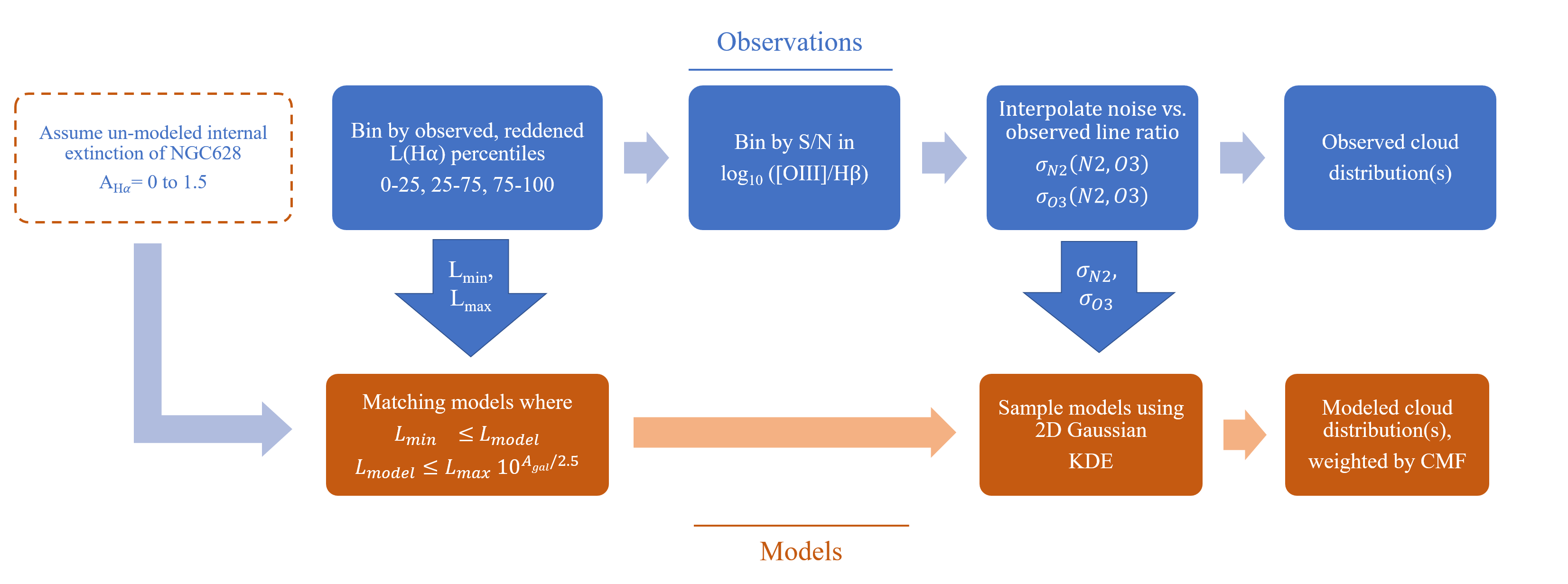}
    \caption{A flow chart representing the important steps taken to either bin and modify models in order to compare meaningfully with observations. Blue arrows indicate where observational properties are used to select models or modify their distribution. These steps are outlined in detail in Section~\ref{sec:obscomp}.}
    \label{fig:obs_model_comp_flow}
\end{figure*}

 \begin{figure*}
	\centering
	\includegraphics[width=0.9\textwidth]{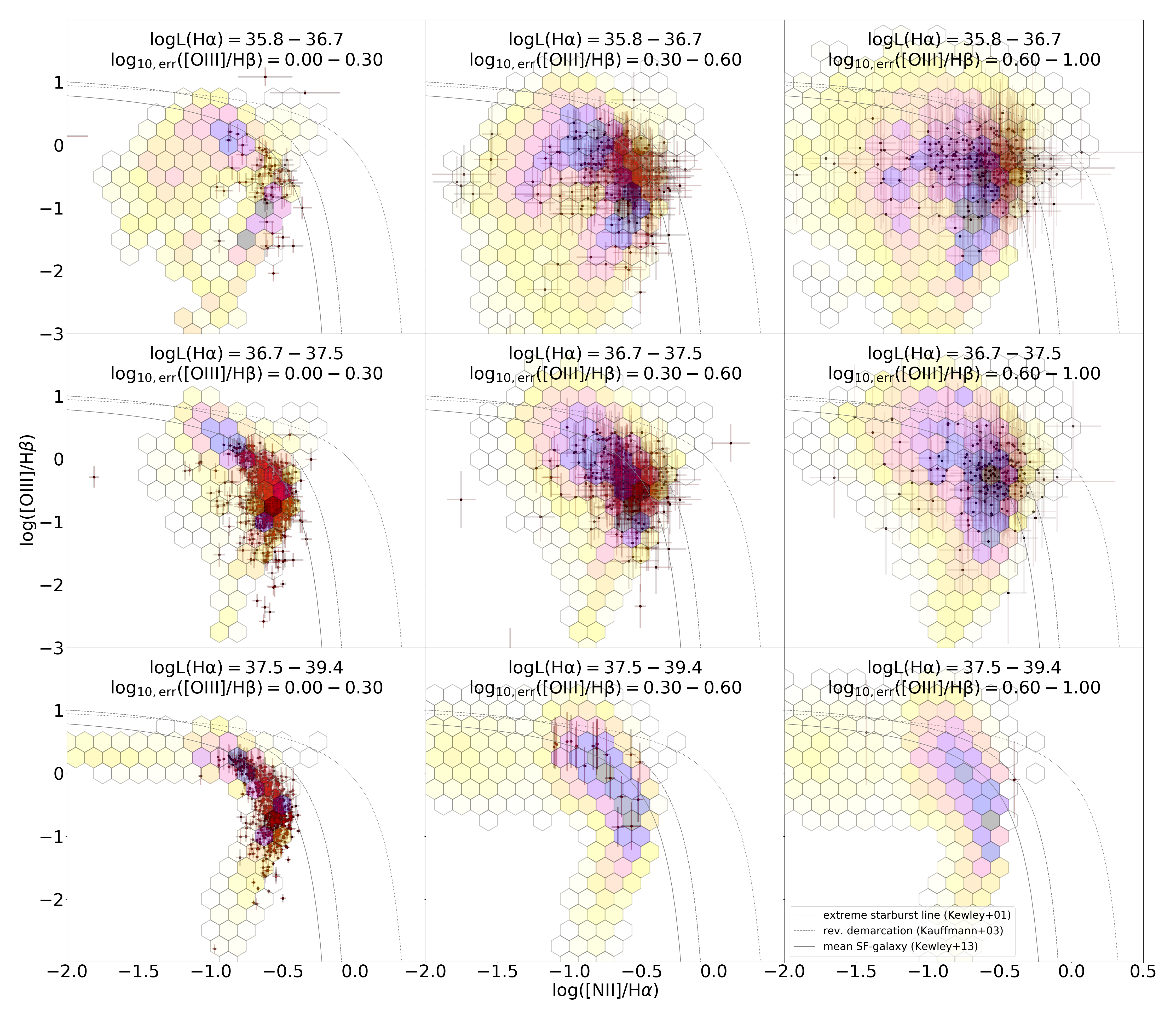}
	\caption[Time weighted BPT diagram]{An example of a comparison between {\sc warpfield-emp} results and real observational data. The color maps show a 2D histogram of the \OIII/\Hb and \NII/\Ha line ratios corresponding to a range of different {\sc warpfield-emp} model outputs. Each one is smoothed using a 2D Gaussian function with dispersions in the cardinal directions computed using the procedure outlined in Section~\ref{sec:obscomp}. This smoothing is designed to be comparable to the errors in the observational data-points. The contributions of the individual model outputs to the 2D histogram are also weighted by a factor that accounts for the cluster mass function and the timestep corresponding to that particular model output. The observational data-points show \OIII/\Hb and \NII/\Ha line ratios for individual \hii regions in NGC~628, taken from  \citep{Rousseau-Nepton2018}. As the {\sc warpfield-emp} models include the effects of dust extinction, we compare them here with observational data that has not been corrected for reddening. To aid in clarity, higher signal-to-noise points have increased opacity. From top to bottom, the panels show samples of \hii regions with increasing \Ha luminosity, corresponding to 0--25\%, 25--75\%, and 75--100\% of the full distribution.  Each column represents a different signal-to-noise interval in \OIII/\Hb, increasing from left to right. The model data shown in each row is selected to have a comparable range of \Ha luminosities to the observations shown in each row, after accounting for the possible effects of foreground extinction.  Finally, we show for reference the \citet{Kewley2001} (dotted line) and \citet{Kauffmann2003} (dashed line) criteria for distinguishing between star-forming regions and AGN, as well as a line showing the mean location of the integrated emission from star-forming galaxies in the diagnostic plot (solid line).}
	\label{fig:N2O3_time_weighted_w_obs}
\end{figure*}

\section{Comparison With Observations} \label{sec:obscomp}
There are two different ways in which we could consider comparing the results of our {\sc warpfield-emp} models with real observational data. The first involves the comparison of models and observations of a single region. For example, in previous work we have found {\sc warpfield} models that are a good match to the current evolutionary state of the \hii regions observed in 30 Doradus in the Large Magellanic Cloud \citep{Rahner2018} and the W49 star-forming region in the Milky Way \citep{Rugel2018}. Using {\sc warpfield-emp}, it is straightforward to make predictions for the line and continuum emission from these regions that can be compared to the observed values. In this case, little post-processing is required to convert the surface brightness maps produced by {\sc warpfield-emp} into a form that can be directly compared with the observations. The maps must be smoothed to match the angular resolution of the observations and noise must be introduced at a level that is consistent with the noise in the real data. As the appropriate values to choose here will be specific to the particular observations of interest, {\sc warpfield-emp} itself does not account for either of these post-processing steps, which must therefore be carried out by the user. This is beyond the scope of the present paper.

The other type of observational comparison is one between a population of objects and a range of different {\sc warpfield-emp} models. This is what we focus on here. It is important, as it tells us how well our models succeed in reproducing the diversity of different \hii regions and wind-driven shells observed in real galaxies. A proper comparison along these lines requires a population synthesis approach to reproduce the observed density of models at each value of \NII/\Ha and \OIII/\Hb ratios, since we cannot assume that we are observing all of the different star-forming regions at the same time in their history. Doing this fully self-consistently lies beyond the scope of the current paper, although we will return to this point in future work. 
However, we can approximate the observed distribution in a diagnostic such as the BPT diagram by appropriate selection and weighting of our model predictions, as we describe in more detail below. 

The set of observational data that we use for the purposes of comparison is a catalog of \hii regions based on tunable filter observations of NGC~628 from \citet{Rousseau-Nepton2018}. That catalog provides line fluxes integrated over the entire extent of each \hii region. We are using a version of the catalog where no reddening correction has been applied, which is significant for comparisons involving \OIII. Direct comparison of our model predictions with a catalog such as this requires some care, as the survey characteristics and catalog classification methods affect the number and properties of the regions included in the catalog. 

To take a simple example, any optically-selected catalog of \hii regions has a clear bias against including young, highly embedded objects, since these will have high extinctions and hence will often have optical line fluxes below the detection limit of the survey. Failing to account properly for effects such as this can therefore lead to apparent discrepancies between models and observations that are purely a consequence of the observational selection effects. 
 
In order to make a meaningful comparison between our set of models and the the \citet{Rousseau-Nepton2018} \hii region catalog we employ the following procedure, outlined in Figure~\ref{fig:obs_model_comp_flow}, to produce Figure~\ref{fig:N2O3_time_weighted_w_obs}. Note that in what follows, we distinguish between a {\sc warpfield-emp} ``model'' (the time history of the emission produced for a single combination of cloud mass, SFE, cloud density etc.) and a ``model output'' (the results of a {\sc warpfield-emp} model at a particular output time). \\ 

\begin{itemize}
\item {\bf Step 1}: \\ 
We produce a list of the \hii regions in the \citet{Rousseau-Nepton2018} catalog ordered by increasing observed \Ha luminosity, L(\Ha). We then divide up the \hii regions from the list into three bins, the first containing the first 25\% of the regions in the list (i.e.\ the 25\% of \hii regions with the lowest observed \Ha luminosity), the second containing the next 50\% and the last containing the final 25\% (row 1, column 1 in Figure~\ref{fig:obs_model_comp_flow}). In practice, this results in a set of bins with boundaries $35.8 < \log_{10} L({\rm H}\alpha) < 36.7$, $36.7 < \log_{10} L({\rm H}\alpha) < 37.5$ and $37.5 < \log_{10} L({\rm H}\alpha) < 39.4$, with \Ha luminosities in erg~s$^{-1}$. Binning the \hii regions in this way allows us to (approximately) delineate regions with different masses and ages. Each bin corresponds to a different row in Figure~\ref{fig:N2O3_time_weighted_w_obs}.
The \hii regions from the \citet{Rousseau-Nepton2018} catalog are reddened not only by dust within the region itself (which is accounted for in our {\sc warpfield-emp} models) but also by foreground dust within NGC~628 and the Milky Way, which is not accounted for in our models. When selecting the set of model outputs to compare to each observational bin, we therefore allow for a range of foreground extinctions $0 < A_{\rm H\alpha, max} < 1.5$~mag (row 2, column 1 in Figure~\ref{fig:obs_model_comp_flow}). Using this range, we identify all model outputs where the predicted emergent \Ha luminosity falls within the range
\begin{equation}
L_{\rm min} < L({\rm H}\alpha)_{\rm model} < L_{\rm max} \times 10^{A_{\rm H\alpha, max}/2.5},
\end{equation}
where $L_{\rm min}$ and $L_{\rm max}$ are the minimum and maximum \Ha luminosity included in the bin. \\

\item {\bf Step 2}: \\
We next divide the observational data into low, medium and high uncertainty subsets, using the size of the 1$\sigma$ error in the value of $\log_{10}$~\OIII/\Hb to distinguish these subsets.\footnote{We use \OIII/\Hb rather than \NII/\Ha to distinguish the different subsets as the observational error in the former is generally higher than that in the latter.} The three subsets correspond to errors in $\log_{10}$~\OIII/\Hb that lie in the ranges 0.0--0.3, 0.3--0.6 and 0.6--1.0, respectively. \hii regions with errors in their measured $\log_{10}$~\OIII/\Hb greater than 1.0 are omitted entirely from the comparison with our models, as the size of their error bars renders any such comparison of very limited value. The three subsets are illustrated in the three columns of Figure~\ref{fig:N2O3_time_weighted_w_obs}. In each panel in Figure~\ref{fig:N2O3_time_weighted_w_obs}, the individual data-points indicate the values of the line ratios for each \hii region located in the corresponding \Ha luminosity and \OIII/\Hb error bin. To aid in the legibility of the figure, we scale the opacity of the error bars with the S/N. Finally, for the \hii regions in each \Ha luminosity and \OIII/\Hb error bin, we perform a 2D bi-variant-spline fit to the the error in \OIII/\Hb and \NII/\Ha as a function of both ratios($\sigma_{N2}$ and $\sigma_{O3}$). This gives us a description of how the error in the line ratios varies as a function of the ratios for the \hii regions in each bin. Note that this approach is more accurate than adopting a single representative error value for each ratio in each bin, as it accounts for the substantial covariance between the errors and the line ratios. \\

\item {\bf Step 3}: \\
At this point, we have nine different sets of observational data-points (corresponding to the different \Ha luminosity and \OIII/\Hb error bins) and three different sets of model outputs (corresponding to the different \Ha luminosity bins, selected as described in Step 1). Each model output yields precise values for the \OIII/\Hb and \NII/\Ha line ratios. However, if we want to compare these values in a meaningful way with the observed values, we need to account for the uncertainty in the observations. Essentially, what we would like to do is to assign each data-point a set of line ratio errors that are comparable to those in the observational data. This is made more complicated by the fact that the size of these errors differs in the different observational subsets and also varies systematically as a function of the line ratios within each subset. Therefore, rather than assigning each model output the same error values, we instead make use of the 2D spline fit described in Step 2. In each \Ha luminosity and \OIII/\Hb error bin, we run through the set of model outputs associated with that bin. For each model output, we determine whether its predicted line ratios lie within 3$\sigma$ of at least one observational data-point. If they do, then we use the spline fit to compute the appropriate errors that we should assign to the model data-point. If they do not -- i.e.\ if the model output lies in a part of the \OIII/\Hb--\NII/\Ha parameter space that is not represented in the observational data -- then we cannot safely use the spline fit. In this case, we instead simply assume that the errors in these model outputs are equal to the median error of the observed points in that particular \Ha luminosity and \OIII/\Hb error bin. Finally, to produce the smooth distribution of simulated data illustrated by the color maps in Figure~\ref{fig:N2O3_time_weighted_w_obs}, we represent each model output data-point and its associated errors as a 2D Gaussian and carry out a weighted sum of these Gaussians.\footnote{In practice, we do this by Monte Carlo sampling the individual Gaussians, but this is an implementation detail that does not significantly affect the form of the final distribution.} As weighting factors, we take the timestep corresponding to that model output\footnote{Recall that {\sc warpfield-emp} employs irregular temporal sampling, depending on how rapidly key physical parameters change, so some model outputs represent a longer time in the history of the system than others. If we were to fail to account for this in our weighting, then we would tend to over-represent results from early in the evolution of the individual systems, when things are changing rapidly, and under-represent results from later in the lifetime of the systems.} in units of $10^{5}$~yr, as well as a factor accounting for the probability of finding a cluster of the particular mass associated with that model output. This latter factor is computed assuming a cluster mass function with a slope of $\alpha=-2.0$. 
 
Note that by construction our set of model outputs only includes results for shells with radii $r \leq 200$~pc, since we assume that on larger scales the effects of galactic shear will tend to disrupt the shell, and hence terminate our {\sc warpfield} calculations for shells that reach this size. We therefore confirm that all of the \hii regions in the utilized observational dataset have sizes smaller than this. In practice, this restriction turns out to be unimportant, as almost all \hii regions observed in NGC~628 have radii $r \ll 200$~pc.

\end{itemize}

Overall there is good agreement between the models and the observations. Most of the observed \hii regions lie on a locus in the \OIII/\Hb--\NII/\Ha plane that is also densely populated by model results, meaning that many of our models spend a large fraction of the time for which they are observable in this region in the plane. Moreover, comparison with the high signal-to-noise data-points shows that the models not only fall in the right location in the plane, they also reproduce the 0.25~dex scatter seen in the \NII/\Ha ratio at constant \OIII/\Hb. Note that for this subset of observational data, the scatter is not simply explainable as a consequence of the statistical errors in \NII/\Ha, as these are small. Interestingly,
we reproduce this scatter despite our models having a constant N abundance, suggesting that it is primarily driven by variations in the temperature and excitation of the \NII-emitting gas, rather than by variations in the metallicity.

Nevertheless, we also see that there are regions in the plane where we have model results that correspond to few or no observed \hii regions. For example, at log~\NII/\Ha $\sim 1$ and high log~\OIII/\Hb ($\ge 0$), we find a clear concentration of model outputs but only a few actual \hii regions. As we discuss in Section~\ref{sec:analysis}, models producing outputs in this part of the parameter space typically have high initial cloud densities, so the lack of observed \hii regions in this area of the plot may simply be telling us that there are few giant molecular clouds (GMCs) with such high initial densities in NGC~628. This would be consistent with the relatively small cloud-scale molecular gas densities that one infers from ALMA observations of GMCs in NGC~628 \citep{Sun2018}, although one should bear in mind that the latter values are likely biased low by the effects of beam dilution. 

The second main region of interest are low ionization objects, with $\log [\ion{O}{III}]/\Hbmath \le -1.0$ and $\log [\ion{N}{II}]/\Hamath \sim -0.6$. A few of our models do in fact populate this part of the observational plane. However, it remains under-sampled as it corresponds to calculations with $M_{\rm cl} \le 10^{5} \, {\rm M_{\odot}}$ and with star cluster masses below a few $1000 \, {\rm M}_{\odot}$. For such small clusters, our assumption of a fully-sampled IMF is no longer valid, and so reproducing the observational data in this region will likely require us to consider models with stochastically-sampled IMFs, a topic which is outside of the scope of this paper.

\section{Analysis}
\label{sec:analysis}

\begin{figure}
    \includegraphics[width=0.92\linewidth]{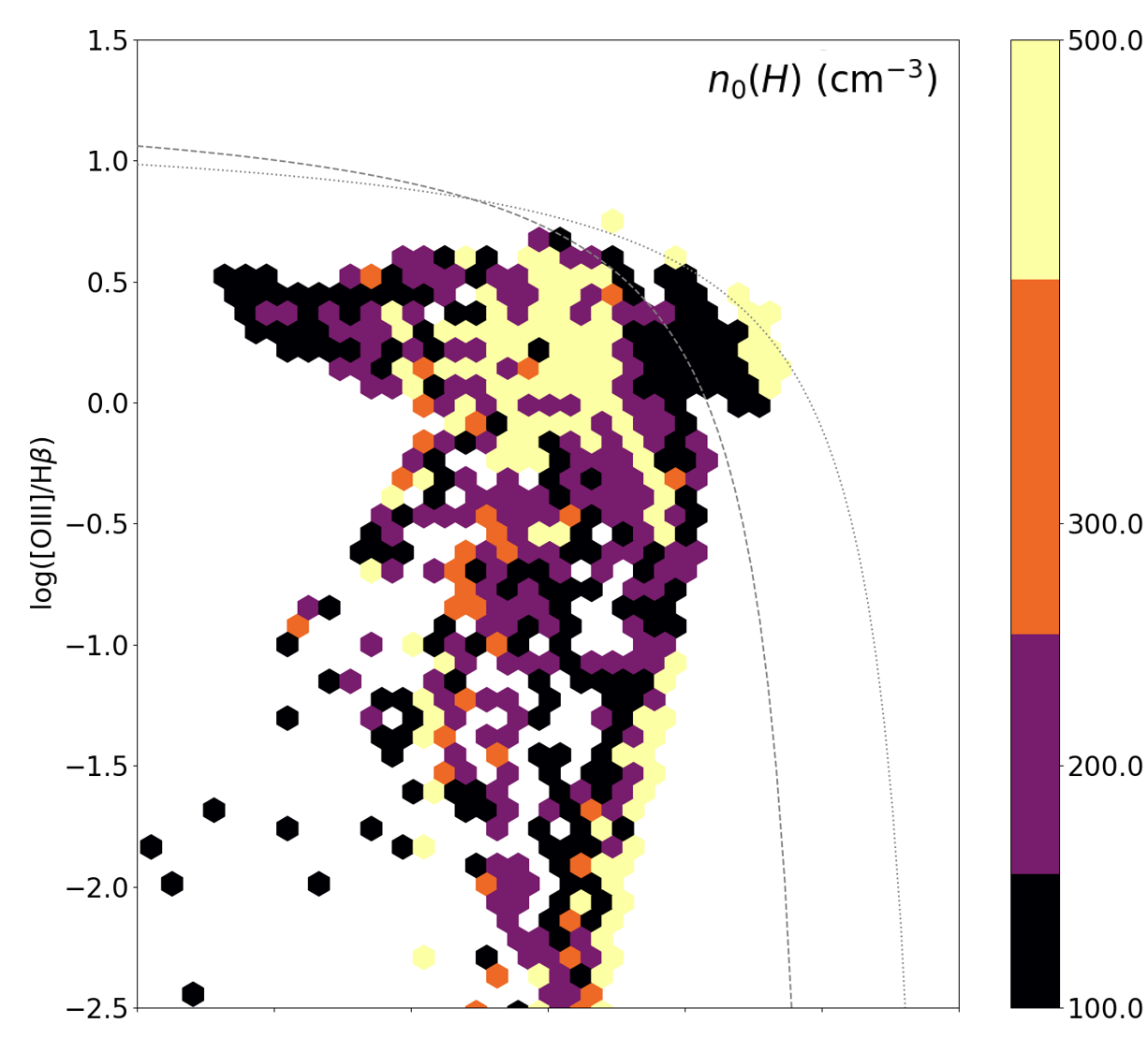}
    \includegraphics[width=0.92\linewidth]{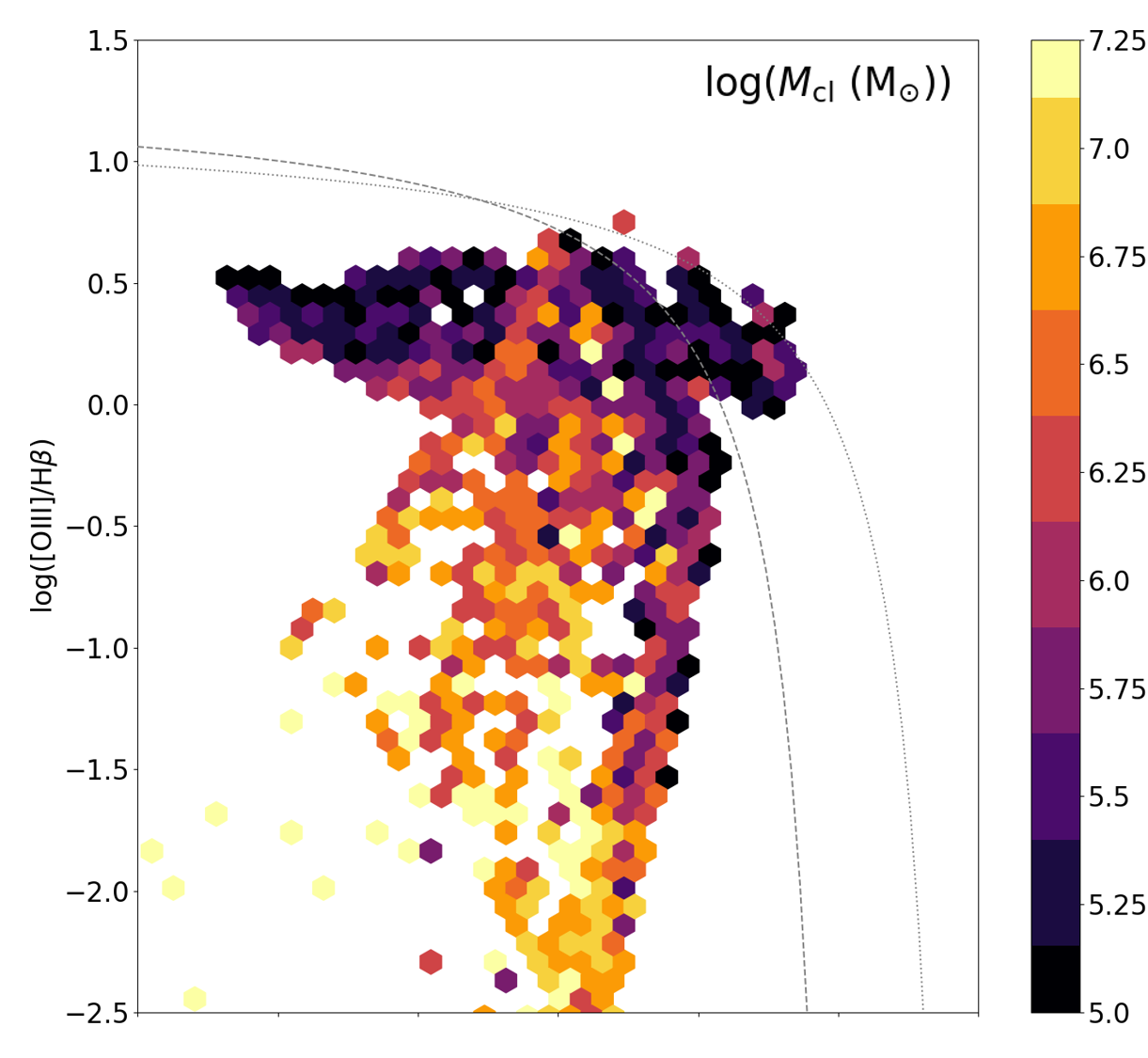}
    \includegraphics[width=0.92\linewidth]{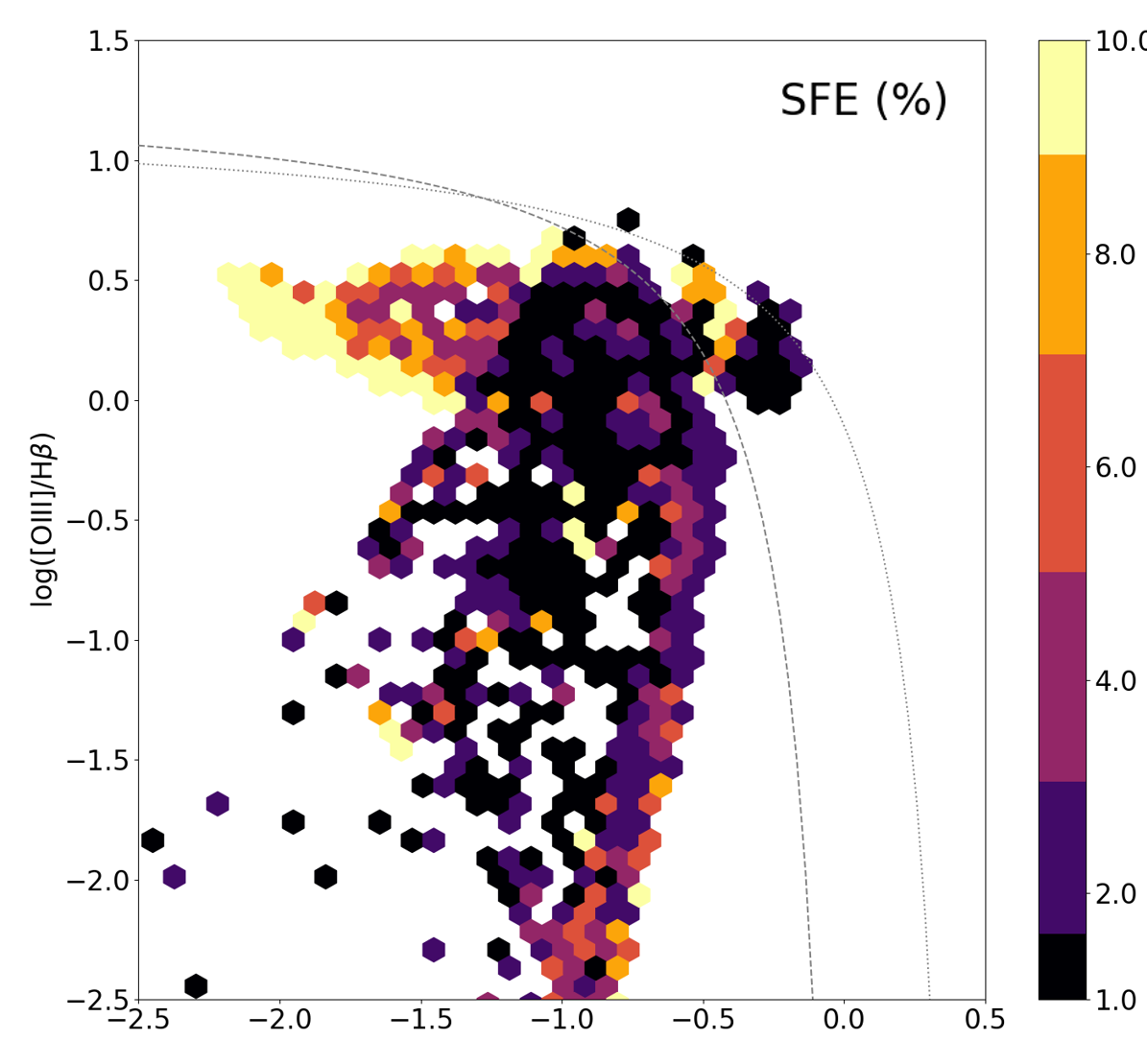}
    \caption{{\it Top:} BPT diagram for our set of \hii region models, colour-coded by the value of the initial cloud density most likely to result in emission at that point in the diagram. Specifically, in each bin we show the time-step weighted mode, i.e.\ the value of $n_{0}$ that produces the set of models that collectively spend the most time in that bin. {\it Middle, bottom:} the same, but for the cloud mass and input SFE, respectively. Note that in all three panels, objects are only included if they have \Ha luminosities exceeding $10^{35.8} \, {\rm erg \, s^{-1}}$ and sizes smaller than 200~pc, as in Section~\ref{sec:obscomp}.}
    \label{fig:BPT_NII_model_parm}
\end{figure}

\subsection{Dependence on Initial Cloud Properties}
\label{sec:dep_initial_cloud_prop}
One interesting application of {\sc warpfield-emp} is to explore whether the location of star clusters in the BPT diagram can tell us anything directly about their properties, or the properties of the cloud in which they formed. To this end, in 
Fig.~\ref{fig:BPT_NII_model_parm} we show where models with different values for the main input parameters (initial cloud density, cloud mass, and input SFE) fall in the BPT diagram. In each hexbin in the plots we present the mode, i.e.\ the most frequent value inside that bin. This should not be confused with the actual observational probability, since our grid does not account for the fact that some input parameters -- e.g.\ lower cloud masses -- occur more often than others. We do, however, account for the use of adaptive time steps by weighting each point in the BPT-diagram by the length of its corresponding time step.\footnote{Roughly speaking, the simulation time step corresponds to the time a model spends inside a bin. This is only true as long as model does not move more than one bin size on the diagnostic diagram during one time step.}

As can be seen in Fig. \ref{fig:BPT_NII_model_parm}, the initial cloud density is not a good indicator of where a model will fall, even after pruning models due to observability. There is a slight tendency for \hii regions born in an initially low density cloud to populate the top left corner of the BPT diagram, i.e.\ low \NII/\Ha and high \OIII/\Hb, while \hii regions born in high density clouds tend to be found in the low \NII/\Ha and low \OIII/\Hb tail of the distribution. 
However, we also see that there are many high density models that spend part of their time in the low \NII/\Ha and high \OIII/\Hb region of the plot, and many low density clouds in the low \NII/\Ha and low \OIII/\Hb tail. This behavior is a consequence of the time dependence of the emission line ratios that we already discussed in Section~\ref{sec:results1}. Because of this time dependence, a given cloud model does not produce line ratios that always remain in the same part of the BPT diagram, but instead yields values that move around over time. 

We find similar behavior if we vary the other two input parameters, cloud mass and SFE. Clouds with high SFEs tend to be found at the top of the diagram, while very massive clouds mainly fall around the inner bend, but in each region of the plot there is also considerable scatter. Therefore, if the only information we have available on a given cluster and cloud is its location in the BPT diagram, we cannot conclusively infer the cloud or cluster mass, or the mean density of the cloud. The best we can do is to make a probabilistic statement, e.g.\ that an \hii region located at the top of the plot is more likely to be located in a low mass cloud than in a high mass cloud. Converting this qualitative statement into a quantitative statement and accounting for the fact that in reality the line ratios for any given \hii region are not known precisely is an important task, but one which lies outside of the scope of this preliminary investigation (see \citealt{Ardizzone2018} for a first assessment).
 
\subsection{Dependence on Evolved Cloud Properties}
\label{sec:dep_evolved_cloud_prop}

\begin{figure*}
    \includegraphics[width=0.9\textwidth]{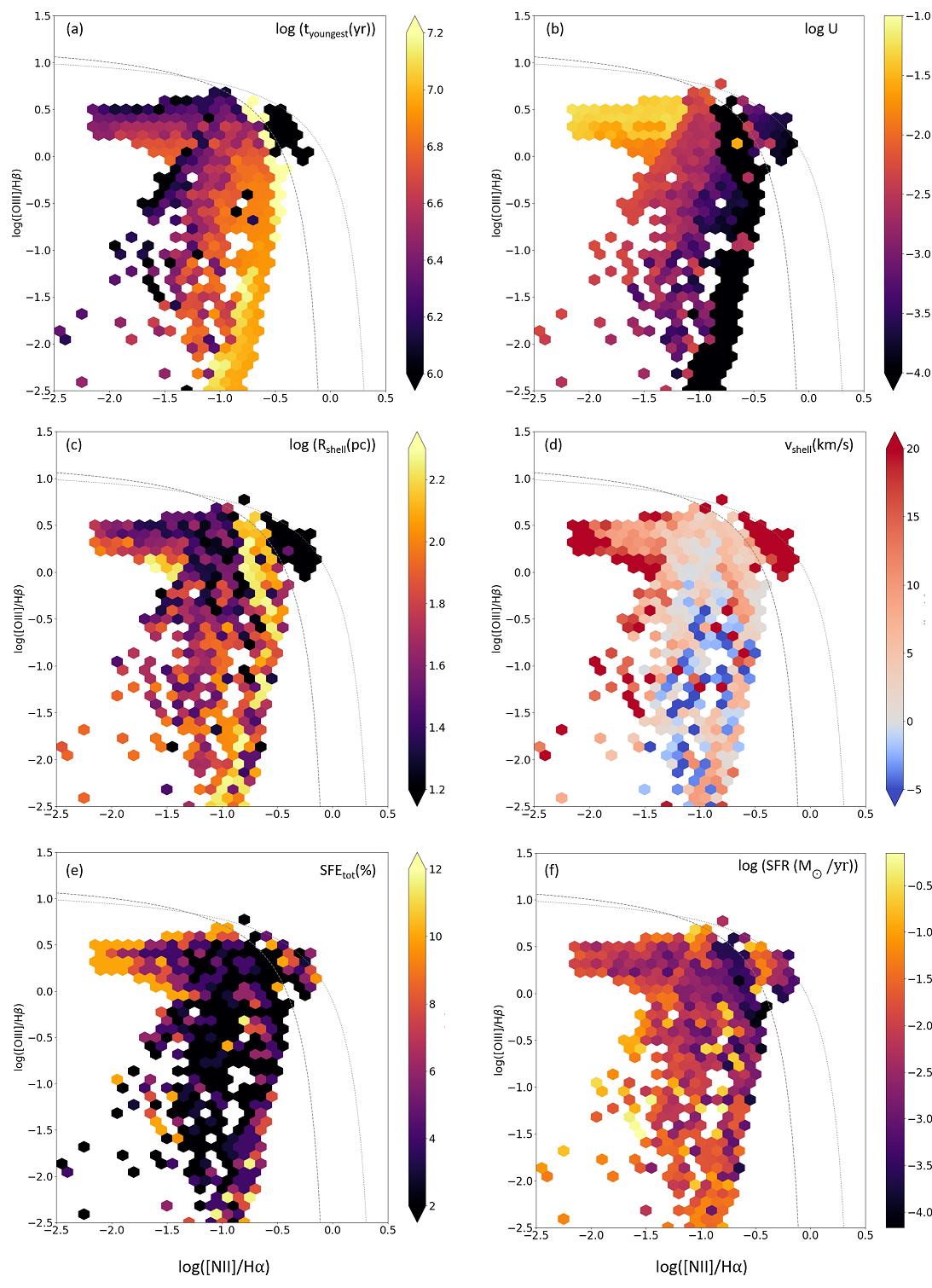}
	\caption{Same as Fig.~\ref{fig:BPT_NII_model_parm} but color coded by instantaneous cloud conditions: (a) the age of the youngest cluster that has formed, (b) the ionization parameter $U$ measured at the inner shell radius, (c) the shell radius, (d) the shell velocity (objects with negative velocities are contracting), (e) the total SFE of all starbursts up to that age, and (f) the derived average star formation rate.}
	\label{fig:BPT_NII_model_derived}
\end{figure*}

As well as looking at the connection (or lack thereof) between the initial conditions for each model and its later location in the BPT diagram, we can also investigate how this position correlates with properties defining the instantaneous conditions within the \hii region, such as the age at which we observe it. In Figure~\ref{fig:BPT_NII_model_derived}, we show BPT diagrams similar to those in Figure~\ref{fig:BPT_NII_model_parm}, but color-coded (a) by the age of the youngest cluster; (b) the ionization parameter, $U$, measured at the inner shell radius; (c) the size and (d) the expansion velocity of the shell; (e) the net star formation efficiency SFE$_{\rm tot}$; and (f) the time-averaged star formation rate (SFR). The net star formation efficiency is defined here as the ratio of the total stellar mass formed within the cloud to the initial cloud mass. In each case, we show the median value in each hexbin. In the cases where the feedback-driven shell never recollapses, the net star formation efficiency plotted in panel (e) is the same as the star formation efficiency \SFE that we specify as an input parameter. However, in cases where the shell collapses one or more times during the lifetime of the cloud, we form additional stars in a burst after each re-collapse, and in this case the net star formation efficiency is larger than the input value. Finally, the time-averaged star formation rate shown here is defined as
\begin{equation}
{\rm SFR} = \sum_{i} M_{{\rm cluster}, i} / {\Delta t_0} 
\end{equation}
where $M_{{\rm cluster}, i}$ is the mass of each individual cluster, with a total number of clusters equal to one more than the number of re-collapses, and $\Delta t_0$ is taken as the total age of the model. Unlike SFRs derived from large ensembles of objects, where it makes statistical sense to average the SFR over a given time frame (often taken to be 10~Myr; \citealt{KennicuttJr.1998}), the values we report here assume that each individual SF event is instantaneous. 

In contrast to Figure~\ref{fig:BPT_NII_model_parm}, several clear correlations are visible in Figure~\ref{fig:BPT_NII_model_derived}. There is an obvious relationship between the age of the youngest cluster, the value of the ionization parameter and the position of the cloud in the BPT diagram, which we discuss in more detail below. There is also a noticeable but somewhat weaker correlation between the size of the shell and the value of the \OIII/\Hb ratio. Clouds with high \OIII/\Hb often have small radii, although there are some models in this regime that have large radii. On the other hand, shells with large radii generally have low \OIII/\Hb.

This is easily understood as an evolutionary effect. The size of the feedback-driven shell is correlated to the age of the youngest cluster. Compact expanding shells are generally found around young clusters, while star clusters whose feedback has driven the shell to radii $\gtrsim 100$\,pc tend to be old. Therefore, objects found in the lower half of the BPT diagram are those with large shell radii. In galaxies with low shear, where shells larger than 200\,pc remain coherent, objects with very large shells would also inhabit the region around the Kauffmann bend but would be hard to observe due to their low surface brightness. Note also that although compact shells can be found around older clusters in cases where the shell is collapsing, these objects are generally not observable. This is clear from panel (d) in Figure~\ref{fig:BPT_NII_model_derived}, where we see that only a few of the \hii regions that pass our observability cuts are collapsing. The majority are expanding, with velocities ranging from a few up to 20~km$\, {\rm s}^{-1}$. It should be noted that some contracting objects are hidden by our choice to present the median in each bin, but even if we were to color-code each cell by its minimum velocity, we would still only be left with a handful of contracting models. We therefore predict that almost all contracting objects are unobservable in the optical lines depicted here due to their high extinction and hence low luminosity.

\section{The failure of simple abundance diagnostics}
\label{sec:simple_diagnostics}
The idea of using emission lines from \hii regions to constrain the gas-phase metallicity of the ISM is a very old one \citep[see e.g.][for an influential early paper]{Searle1971}. The current ``gold standard'' methods for doing so involve the measurement of the electron temperature $T_{\rm e}$ using line ratios such as the ratio of the \OIII 4363$\,$\AA~auroral line to the \OIII 5007$\,$\AA~nebular line that are highly sensitive to the electron temperature. Unfortunately, the auroral lines on which these methods are based are weak and are often not detected in observations of extragalactic \hii regions. Consequently, there is ongoing interest in using the ratio of strong lines, such as \NII/\Ha, as metallicity tracers \citep[see e.g.][]{PP2004}.

A basic assumption underlying this approach is that variations in the line ratio being used as a metallicity indicator are driven primarily by variations in the relative gas phase abundances of the elements, or in other words that one can cleanly separate the effects of changes in the illuminating radiation field from changes in the elemental abundances. In the case of the \NII/\Ha ratio, this is often justified by the argument that the similar ionization potentials of N and H make the ratio of N$^+$/H$^+$ insensitive to secondary parameters. What these simple assumptions do not account for is the covariance of many secondary parameters in a way that causes significant changes in the \NII/\Ha ratio at fixed metal abundance.
An in-depth examination of the extent to which this covariance confounds commonly-used abundance diagnostics is beyond the scope of our current paper. Nevertheless, the results presented here are already enough to allow us to draw some simple qualitative conclusions. 

To begin with, if we look at Figure~\ref{fig:BPT_NII_model_derived}, several points become clear right away. First, there is a good correlation between the age of the youngest cluster in the cloud and the position of the cloud in the BPT diagram. Clouds with young clusters (corresponding to high ionization parameters at the inner shell boundary) have high values of \OIII/\Hb whereas clouds with older clusters show a strong decrease in \OIII/\Hb. However, clusters older than $\sim 20\,$Myr, also populate the region close to the bend in the Kauffmann demarcation line (around log~\OIII/\Hb~$\approx 0.2$, log~\NII/\Ha~$\approx -0.5$), but are too faint to be observable and are thus not shown in Fig. \ref{fig:BPT_NII_model_derived}. We also see that there is a very good correlation between the ionization parameter itself and the location of the H{\sc ii} region in the BPT diagram. This result is unsurprising: we know already from static models that there should be a close link between the ionization parameter describing a given H{\sc ii} region and the line emission from that region. However, our models demonstrate that there is a significant degree of scatter in this relationship, particularly at high \OIII/\Hb. Observationally, when such scatter is seen, it is often interpreted as a metallicity effect, as changes in the metallicity will generally have the effect of changing \NII/\Ha while leaving \OIII/\Hb approximately constant. However, all of our models 
were calculated at the same metallicity, and so our results clearly demonstrate that metallicity variations are not the only possible cause of the observed scatter.

The top two panels in Figure~\ref{fig:BPT_NII_model_derived} also illustrate another important result. They demonstrate that there is a good correlation between the value of the ionization parameter at the inner edge of the H{\sc ii} region and the age of the youngest cluster within the H{\sc ii} region. Since the cluster age controls the hardness of its spectrum (see Section~\ref{sec:starcluster}), another way of stating this is that the value of ionization parameter is correlated with the hardness of the spectrum. Therefore, we cannot vary these quantities independently when modelling real H{\sc ii} regions \citep[see also][]{Dopita2006}.

\section{Embedded objects}
\label{sec:embedded_objects}

As we have shown in \citet{Rahner2017a}, the minimum star formation efficiency necessary in order for stellar feedback to be able to disrupt a cloud is a function of the cloud mass, metallicity and natal/ambient density. One reason for this is that for a stellar cluster with a fully sampled IMF, the energy and momentum available to drive feedback scale linearly with the mass of the cluster, as $M_{*} = \epsilon_{\rm SF} M_{\rm cl}$, but the binding energy of the cloud scales as $M_{\rm cl}^{2}$. Therefore, the amount of feedback per unit of binding energy is not constant, unlike the mass to light ratio of the cluster or the feedback per unit mass ratio, which are both constant for constant \SFE. Instead, the feedback per unit of binding energy decreases as $M_{\rm cl}^{-1}$, leading to larger clouds being more resistant to feedback that smaller clouds. 

Furthermore, the natal cloud density also plays an important role. The gravitational binding energy at constant mass is inversely proportional to density, implying that, everything else being equal, denser clouds are harder to destroy. Denser clouds also cool faster, which reduces the effectiveness of stellar wind feedback at early times. On the other hand, they can also be more effective at capturing the full radiation output of the stellar cluster, and so radiation pressure can be {\em more} effective in denser clouds. Finally, varying the metallicity also affects the relative importance of winds and radiation pressure. Altogether, the overall impact of feedback depends on the parameters of the natal cloud and the central cluster in a highly complex and non-linear fashion \citep{Rahner2017a,Rahner2019}.

In practice, we find when we vary these parameters that star clusters and their parental clouds form two distinct populations of star-forming regions, with significant differences in both their observational appearance and their star formation histories. We refer to these as unembedded and long-term embedded clusters. Their primary distinction lies in the observability of their stellar clusters. A common observational definition used to distinguish between embedded and unembedded clusters is the mean visual extinction to the cluster: 
clusters with $A_{\rm V} \geq 5.0$ are embedded, while those with $A_{\rm V} < 5.0$ are unembedded \citep{LadaLada2003}. We discuss the properties of these two populations in more detail below.

\subsection{Unembedded clusters}
All our natal clusters are initially ``embedded" by the $A_{\rm V}$ criterion mentioned above, as expected, but in many cases feedback-driven dynamical evolution rapidly decreases the extinction to the point where we would classify them as unembedded. This is typically associated with the expansion of the shell radius beyond the initial cloud radius, as beyond this point further expansion sweeps up little additional mass from the diffuse ISM but continues to increase the surface area of the shell, resulting in a steady decrease in the column density. 

These unembedded clusters have low $A_{\rm V}$ (by definition) and are readily observable because they are still young and hence retain most of the massive stars responsible for ionizing the surrounding gas. The common optical lines used to trace \hii regions (H$\alpha$, H$\beta$, \OIII, \NII, etc.) are therefore bright. Typically, these models correspond to lower mass star-forming regions with moderate densities (100--200~cm$^{-3}$) that undergo one or two bursts of star formation before successfully disrupting the surrounding cloud and becoming unembedded. Broadly, these regions in which expansion and cloud disruption were successful are the ones explored in most optical surveys of \hii regions.

\subsection{Embedded clusters}
In contrast to the above, we can also identify a class of objects characterized by a long-duration embedded phase, one in which feedback is initially unable to disrupt the cloud and star formation continues over an extended period. This behavior is more likely with more massive clusters, owing to the decrease in the amount of feedback per unit binding energy discussed above. This stands in contrast to the simple picture in which it is often assumed that larger objects lead to monotonically increasing expansion  and eventual cloud disruption \citep[see e.g.][]{Murray2011}. Others have also modeled the clouds in massive clusters \citep{Silich2013,Silich2017}, including early predictions of failed feedback due to winds \citep{Silich2004a}. Advancements were made by including the effect of radiation, but these neglected gravity, or failed to treat radiative coupling in clouds of finite mass with self-consistently evolving stellar populations \citep{Martinez-Gonzalez2014}. When considering these effects we find clouds will often not reach Phase III in our evolutionary model (the phase in which the entire cloud is swept up), or if they do, they will undergo re-collapse before expanding far. 

We can easily identify these objects from within our set of {\sc warpfield} models because a significant fraction of massive stars leave the main sequence and die while in the embedded phase (our Phase I and II). Observationally, they are poorly traced by the optical diagnostics that are the main focus of the examples presented in this paper, but they are bright at mid-infrared (MIR) and radio wavelengths. Since {\sc warpfield-emp} also models the emission at these wavelengths, we can compare the predictions of our models with the results of MIR or longer wavelength surveys, although actually doing so lies outside of the scope of our current paper.

Finally, we note that the existence of embedded clusters that have undergone re-collapse is in itself an important prediction of our model, since it implies that embedded clusters need not be young. Clusters in which feedback has failed to disperse the gas may continue to form stars at a slow rate for an extended period, but may be misidentified as massive young protoclusters. 
\section{The Nature of Feedback-regulated line emission} 
\label{sec:nature}
A natural question that the reader may have at this point is why the approach to modelling \hii region and PDR diagnostics outlined in this paper represents an improvement over the more conventional method of scanning large grids of static models computed by individually varying parameters such as the ionization parameter, the gas density or metallicity, or the age of the central stellar cluster. The central motivation for our dynamical approach is the existence of strong correlations between these different parameters. These correlations mean that in reality we are not free to vary the main parameters controlling the emission independently. For example, the density of the ionized shell is set by the boundary pressure exerted on it by the combined stellar wind from the cluster, which in turn depends on the age, mass and metallicity of the cluster, the current size of the shell, and the phase of the evolution of the region (i.e.\ whether or not it still retains its hot gas). Therefore, in order for our \hii regions models to be self-consistent, we cannot simply vary the density without also varying some or all of these other parameters. The advantage of {\sc warpfield-emp} over more static approaches is that it provides us with a simple way of determining which sets of physical parameters are consistent, i.e.\ which parts of the large parameter space are actually touched upon during the evolution of the feedback-driven shells and which are not. 
As examples of the power of this approach, we discuss in
Section~\ref{sec:mean} the implications of our results for observations of diagnostic lines on the scale of entire galaxies (i.e.\ observations that do not resolve the individual \hii regions), and in Section~\ref{sec:SFE} what {\sc warpfield-emp} can tell us about the range of cloud-scale star formation efficiencies that we are likely to encounter in reality and what this implies regarding their location in the BPT diagram.

In addition, {\sc warpfield-emp} also provides us with a means for exploring the impact of particular feedback processes on observable diagnostics of \hii regions since we know at any given time in the models not only how significant the process is at that particular moment, but also how significant it was in the past. This allows us to draw some important conclusions regarding the role played by winds and SNe, which we explore in Sections~\ref{sec:winds} and \ref{sec:SNE} below. 

\subsection{Implications for Mean Galaxy Spectra}
\label{sec:mean}
Here we note that when weighted by time the distribution of objects tends more toward the mean star-forming galaxy track of \citet{Kewley2013}. This  effect is particularly apparent when we look at the brightest quartile of the \hii regions (lowest row in Figure~\ref{fig:N2O3_time_weighted_w_obs}), which are the objects that generally dominate the mean galaxy spectrum. This is noteworthy, given the wide distribution of the underlying predicted observations. It is a consequence of the self-regulation of star formation in our model clouds. This leads to a covariance between the cluster and cloud parameters which heavily favors observations consistent with this track. A further implication is that this trend is less sensitive to the initial cloud properties, and the emission properties are thus a result of the nature of feedback physics, and the metal abundance of the ISM (which would shift the position of the distribution).

\subsection{Star Formation Efficiency} 
\label{sec:SFE}
In our {\sc warpfield} models, the star formation efficiency associated with a given burst of star formation is a free parameter. However, because our models account for re-collapse, the total star formation efficiency -- i.e.\ the total mass of stars formed by the end of the life of the cloud -- is not completely free. Clouds with very low star formation efficiencies per burst undergo repeated cycles of re-collapse and star formation until they reach a point at which feedback is finally able to disperse them, preventing any further star formation. As a result, our models predict that clouds should have a minimum total star formation efficiency that depends primarily on their surface density, with a value of a few percent for surface densities comparable to local GMCs, increasing to 5-10\% in higher surface density clouds \citep{Rahner2019}. Moreover, although our models do not constrain the maximum star formation efficiency of the clouds, it is reasonable to assume that this will not be much larger than the minimum value required to destroy the cloud.\footnote{For this assumption to be invalid, the timescale for star formation in the cloud would have to be much shorter than the timescale on which feedback disperses the cloud. This is true by construction in our models (since we assume that stars form instantly), and would be relatively easy to arrange if supernovae were the main source of feedback, given the time delay between the onset of star formation and the explosion of the first supernovae. However, in practice, winds and radiation play a dominant role at early times \citep[see e.g.][]{Peters2016,Rahner2017a,Rahner2019,Kruijssen2019} and begin to affect the cloud only a short time after the formation of the first massive stars.}

\begin{figure}
	\centering
	\includegraphics[width=0.45\textwidth]{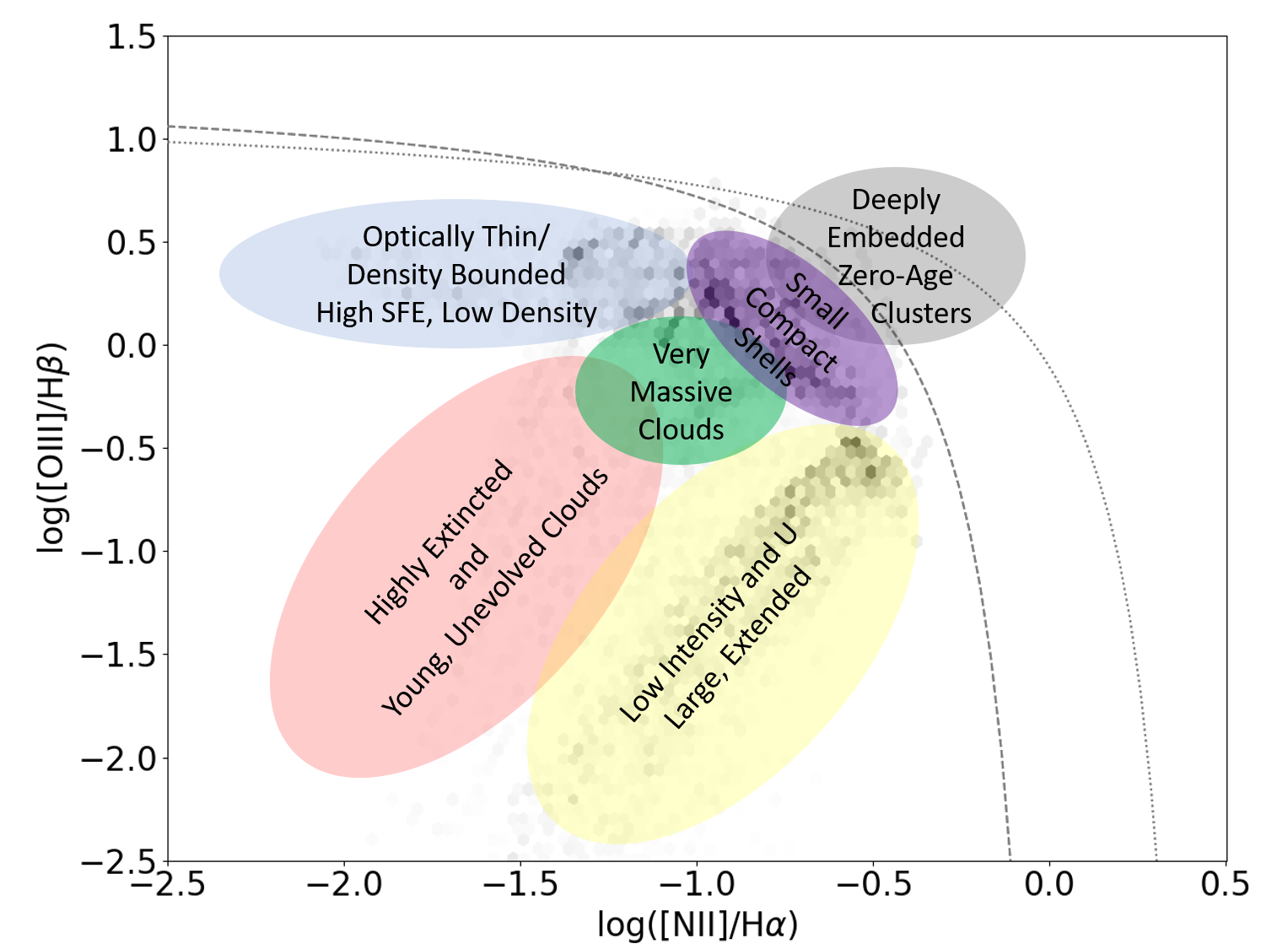}
	\caption[Cartoon Area]{A generalization of where different classes of objects lie in the observable space. High star formation efficiency, high feedback-to-mass ratio objects (blue) and compact \hii regions and clouds (purple) populate the high \OIII/\Hb range. Very massive clouds (green) ($M_{\rm cl} > 10^6 M_{\odot}$), in which the shell remains highly pressured for an extended period, lie somewhat below the compact regions and tend to have suppressed \NII/\Ha. Below this, at $\OIII/\Hbmath \leq -1.0$,
	we find highly extincted \hii regions (red), which also have very low \NII/\Ha, and extended, old \hii regions that have low ionization parameters (yellow). In dark grey we highlight a region populated by massive, young star-forming regions, with hot interior bubbles. These are deeply embedded super star clusters, and within the realm of observability for deep optical surveys, although distinguishing them from the non-stellar sources that also populate this part of the BPT diagram may prove to be a challenge.}
	\label{fig:Cartoon2}
\end{figure}

Despite the inherent degeneracy in the limited set of observational diagnostics considered in this paper, our approach leads to some clear predictions about where particular types of cloud should fall in the BPT diagram. In Figure~\ref{fig:Cartoon2}, we highlight regions in this diagram populated by unique classes of objects. Objects that have high \OIII/\Hb but low \NII/\Ha (blue) correspond to strong starbursts combined with rapidly expanding shells. As Figure~\ref{fig:BPT_NII_model_derived} makes clear, we expect to find this combination primarily in clouds with high star formation efficiencies. If the simple picture we have sketched above is correct, and clouds rarely develop star formation efficiencies much greater than the minimum amount required for cloud destruction, then these objects should be relatively rare in normal spiral galaxies, being confined primarily to regions with high surface density. 

The other class of objects that we expect to have high \OIII/\Hb ratios are small \hii regions exposed to feedback from young clusters (purple). These objects have higher \NII/\Ha ratios than the strong starbursts and have high \OIII/\Hb primarily because of their small size and young age. They need not have high star formation efficiencies.  

In Figure~\ref{fig:BPT_NII_model_parm} very massive clouds with $M_{\rm cl} > 10^{6} \: {\rm M_{\odot}}$ (green) lie somewhat below the small \hii regions in the BPT diagram. They maintain relatively high \OIII/\Hb values for an extended period, because of the time it takes to completely sweep up the gas in these clouds. Below these lie highly extincted regions, which have low values for both line ratios (red) and extended, old \hii regions with low ionization parameters (yellow), which have low \OIII/\Hb but somewhat larger \NII/\Ha. As we have already discussed (see Figures~\ref{fig:BPT_time_evol} and \ref{fig:N2O3_time_weighted_w_obs}), both of these types of object may be difficult to observe in extra-galactic surveys, particularly if one requires detections of both the \OIII and \NII lines in order to classify them as \hii regions, or when their internal attenuation causes them to become as faint as the diffuse ionized gas (DIG) \citep{Pellegrini2019a}.

\subsection{Significance of Winds}
\label{sec:winds}
Winds provide an important boundary pressure that plays a major role in determining the density of the feedback-driven shell. This in turn has a substantial influence on how well radiation can couple to the shell. Therefore, even when winds are not a dominant source of feedback in terms of their energy or momentum, they still play a large role in controlling the effectiveness of radiative feedback. Simulations without winds will generally not produce the correct density structure in the gas surrounding the cluster and hence will not correctly capture the effects of radiation pressure.

In terms of observable diagnostics, our models show that the details of the emission produced by the shell are sensitive to the thickness of the shell and hence to the ratio of the wind boundary pressure to the radiation pressure. This changes over time as the cluster ages and the wind-blow bubble cools and expands. 

Even before we explore the impact of changing metal abundances, and the change in wind strength that follows from this, it is clear that the change in the relative thickness of the \hii region due to the evolution of the cluster and the expansion of the bubble is a detail that cannot be ignored, as it can result in orders of magnitude changes in the relative strengths of low and high ionization lines.

\subsection{The importance, or lack thereof, of SNe}
\label{sec:SNE}
An unexpected result of our models is our finding that many of the evolving clouds, particularly the least massive ones, meet our dissolution criteria immediately after SNe start exploding, but that the SNe themselves are not responsible for destroying the cloud. Recall from Section~\ref{sec:wf} that our primary criterion for cloud dissolution is that the densest portion of the cloud must have a density $n < 1 \: {\rm cm^{-3}}$ for at least 1~Myr. In the clouds in question, the feedback-driven shell sweeps up the entire cloud before any of the massive stars in the central cluster reach the end of their lives (i.e.\ the clouds reach Phase III in the terminology of Section~\ref{sec:EMP}). While all of the stars in the cluster remain alive, the density of the gas in the shell remains above $1 \: {\rm cm^{-3}}$ owing to the confining pressure of the winds and radiation. However, once some of these stars start to explode as SNe, the momentum available in the winds and the radiation that couples to the shell rapidly decreases, despite the positive contribution made by the SNe themselves. As a result, the interior pressure of the cloud drops quickly, resulting in the peak density falling below our dissolution criterion.

Clouds that behave in this fashion will tend to leave behind relatively bright clusters which continue to ionize their surroundings. Moreover, these clusters will be associated with evidence of recent SN explosions. Observationally, one might therefore incorrectly conclude due to the timing that it is the SNe that are responsible for the destruction of the cloud, when in reality, it is the drop in internal pressure due to the aging of the cluster that leads to the shell density dropping and the shell becoming difficult to observe (or indistinguishable from diffuse ionized gas).

\section{Summary}
\label{sec:summary}
In this paper, we have presented the {\sc warpfield-emp} framework, which is comprised of {\sc warpfield}, which we use to model the dynamics of feedback-affected gas around star clusters, the {\sc Cloudy} spectral synthesis code, which we use to model the line and continuum emission produced by this gas and its associated dust, and {\sc polaris}, which we use to account for the effects of dust extinction and line opacity on the emergent radiation. 

{\sc warpfield-emp} allows us to model the time evolution of feedback in molecular clouds without ignoring any of the physical processes that are important for regulating emission from the clouds. The list of physical processes included in the model includes stellar winds, radiation, supernovae, gravity, thermal conduction, cooling of the hot wind-blown bubbles directly via radiation, and indirectly by the introduction of cold entrained shell material. All of these processes are interconnected and highly non-linear, and hence cannot be studied reliably in piecemeal fashion. 

The main approximation made in the {\sc warpfield-emp} approach is the assumption of 1D spherical symmetry (although the model includes non-symmetric effects such as shell fragmentation in a subgrid fashion), a common assumption in nearly all extra-galactic photoionization modeling. Making this approximation reduces the computational cost of the model to the point where it becomes plausible to study large samples of clouds, something which will remain out of reach in 3D simulations for the forseeable future. {\sc warpfield-emp} allows us to study large parts of the cloud and cluster parameter space. And so it is no longer easily justifiable to treat aspects of \hii region physics such as the ionizing SED, ionized gas pressure and density, or the size of the region as independent parameters, nor to ignore the extensive properties of \hii regions and PDRs, such as observed luminosity and cloud mass. As we have shown in this paper, these parameters are not independent and ignoring the substantial covariance between them can easily lead to erroneous conclusions.

We have explored a large parameter space spanning low to high cloud and cluster masses and a range of different cloud densities as a first step toward understanding how nature draws from the cloud and cluster mass function and convolves that into observable diagnostics. We have compared the predicted emission from our model clouds to BPT diagrams of \hii regions in NGC~628 as a first step toward verifying the approach. We find very good agreement between our models and the observational data once we properly account for the observational selection effects. Notably, we reproduce not only the shape of the locus of observed \hii regions in the BPT diagram but also the scatter. This scatter has often been attributed to metallicity variations, but our constant metallicity models demonstrate that it may simply reflect differences between the evolutionary state of different \hii regions.

In a pair of companion papers (including \citealt{Pellegrini2019a}) that use our {\sc warpfield-emp} framework as the basis for a sophisticated population synthesis treatment of galactic emission, we show that the models also do a good job of reproducing the intensity and structure observed in all-sky maps of Galactic \Ha emission as well as measurements of Faraday rotation.

Our models allow us to predict the emission from a large population of evolving clouds in a fully self-consistent fashion. Despite the inherent degeneracy in the diagnostics considered in this paper, they allow us to draw a number of important conclusions: 
\begin{enumerate}
	\item {\bf There is a causal relationship between the feedback, and cluster/cloud evolution which determines the ionizing SED, ISM pressure, ionization parameter and cloud structure.} 
	
The coupling of these parameters is set by a combination of microphysics (which can be modelled from first principles) and the cumulative effect of feedback on the global dynamics of the clouds. Together, these determine both the time evolution of the cloud properties and the star formation history of the clouds. \\

	\item {\bf Constraining photoionization model parameter spaces with dynamics breaks key degeneracies in very simple models between age, size and luminosity.} 

	Replacing the old way of modeling nebulae \citep[e.g.\ by computing static grids of quantities such as ionization parameter, metal abundance, stellar effective temperature, etc.; see e.g.\ ][]{Kewley2001, Kewley2002, Kewley2013, Pellegrini2011, Byler2017} with actual clouds that evolve self-consistently breaks some important degeneracies (although it also introduces new ones related to natal cloud properties). For expected cloud properties, even at constant metal abundance, photoionization tracks of realistic objects are not clean, as observed in the common \NII/\Ha vs.\ \OIII/\Hb plane. \\

	\item {\bf In massive clouds and high pressure regions, collapsing objects are often unobservable at optical wavelengths owing to their high extinctions.}
	
	Surveys which identify \hii regions using common optical lines (e.g.\ \Ha) will tend to miss regions undergoing re-collapse, an observational bias which is important to account for when using \hii region properties to derive feedback properties.
	A corollary of this is that embedded clusters observed at mid- or far-infrared wavelengths need not be young, but could in some cases contain stellar populations with ages up to 30 Myr. \\

    \item {\bf Deeply embedded super star clusters appear as faint AGN.}
    
    The location of deeply embedded massive clusters in the BPT diagnostic (see Figure~\ref{fig:Cartoon2}) presents challenging issues to observers who wish for a simple diagnostic to separate active galactic nuclei (AGN) from star formation activity. As shown, the large amount of differential extinction, combined with large internal gradients in ionization parameter can result in star-forming clouds having diagnostics consistent with faint AGN, leading to an unknown number of missing/or wrongly classified super star clusters (SSCs).\\

	\item {\bf Simple line diagnostics, like [N$\,$II]/\Ha, have substantial scatter at constant ionization parameter and abundance.}
	
The primary cause of this scatter is changes in the boundary pressure between different regions (or within the same \hii region at different times), which have a substantial effect on the thickness and density of the ionized shell. Pressure evolution causes \hii regions in clouds of finite mass to migrate between being density-bounded and being radiation-bounded (i.e.\ between having a low or a high optical depth to ionizing radiation).\\

	\item {\bf Clouds with low star formation efficiencies may undergo multiple bursts of star formation.}

If the star formation efficiency is too low, feedback from the star cluster may be unable to unbind the cloud, which will eventually re-collapse and form new stars \citep[see also][]{Rahner2019}. This implies that some ``rejuvenated'' clusters will contain multiple stellar populations with ages separated by 1--10~Myr. A striking example of this may be 30~Doradus in the Large Magellanic Cloud \citep[see e.g.][]{Rahner2018}. Below $1~\mu$m, the emission from these rejuvenated clusters is dominated by their youngest population, hiding older generations of stars in optical and UV observations. This has serious but largely unexplored implications when measuring star formation rates and efficiencies from clusters of multiple ages, with mass traced by different methods (e.g.\ \Ha vs V- or B-band magnitudes).

\end{enumerate}

\section*{Acknowledgements}
SCOG, RSK, EP, DR acknowledge support from the {\em Deutsche Forschungsgemeinschaft} via the Collaborative Research Center (SFB 881) ``The Milky Way System'' (subprojects B1, B2, and B8) and from the Heidelberg cluster of excellence EXC 2181 ``STRUCTURES: A unifying approach to emergent phenomena in the physical world, mathematics, and complex data'' funded by the German Excellence Strategy. 

\appendix

\section{Differences between Cloudy and WARPFIELD}
\label{appendix:diffs}
When calculating the dynamical evolution of a shell with {\sc warpfield} and the detailed chemical composition of the moving shell and static cloud, both {\sc warpfield} and {\sc Cloudy} require boundary conditions for each component. Nominally these are set by the integrated mass. In {\sc warpfield}, the mass of the shell calculated in the thin shell limit is
\begin{equation}\label{eq:Msh1}
M_{\rm sh} = 4 \pi \int_{0}^{r_{\rm b}} \mu \, n_{\rm cl}(r) r^2 dr,
\end{equation}
where $n_{\rm cl}$ is the initial cloud density, $\mu$ is the mean molecular weight of each particle and $r_{\rm b}$ is the size of wind bubble interior to the shell. With a given equation of state setting the density profile, the mass of the shell is also given by
\begin{equation}\label{eq:Msh2}
M_{\rm sh} = 4\pi \int_{r_{\rm b}}^{r_{\rm static}} \mu \, n_{\rm sh}(r) r^2 dr,
\end{equation}
where $r_{\rm static}$ is the boundary between the static cloud and the shell. This value is not known {\em a priori}, but once the shell density structure is known, it can be determined by requiring that Eq.~\ref{eq:Msh2} yield the same shell mass as Eq.~\ref{eq:Msh1}. The remaining static cloud, if it still exists and the shell has not yet entered Phase III, has a mass
\begin{equation}\label{eq:Mst1}
M_{\rm static} = M_{\rm cl} - M_{\rm sh} = 4\pi \int_{r_{\rm static}}^{r_{\rm cl}} \mu \, n_{\rm cl}(r) r^2 dr,
\end{equation}
where $r_{\rm cl}$ is the cloud radius.

{\sc Cloudy} and {\sc warpfield} both assume that heating and cooling are balanced within the shell, and both adopt the same equation of state which defines how the gas density and thermal pressure vary as we change the thermal and/or ram pressure of the wind, the radiation pressure, and the temperature and chemical composition of the gas. However, {\sc Cloudy}'s more detailed treatment of the chemical composition and heating and cooling result in a slightly different density and temperature structure. Consequently, if we were to adopt the same inner and outer radii for the shell in {\sc Cloudy} as in the {\sc warpfield} calculation, we would effectively be using a different shell mass than that in the {\sc warpfield} model. To avoid this problem, we instead recompute the value of $r_{\rm static}$ within {\sc Cloudy}, with the constraint that the shell mass should equal the value from Eq.~\ref{eq:Msh1}. 

\section{Magnetic and turbulent support of the shell}
\label{appendix:Bfield}
In the example {\sc warpfield-emp} calculations presented in this paper, we assume for simplicity that there is no magnetic field and no turbulent pressure support of the shell. However, {\sc warpfield-emp} does offer the user the option of including both effects, and so in this Appendix, we briefly outline how this is done.  

We consider first the case of the magnetic field. We begin by making the important simplifying assumption that any magnetic field that is present in the gas does not directly affect the dynamical evolution of the shell, which continues to be governed by the purely hydrodynamical equations presented in \citet{Rahner2017a}. Our argument for excluding the B-field from the dynamical evolution was already given in \citet{Rahner2017a} and runs as follows. If the magnetic flux is frozen in the gas (i.e.\ if non-ideal MHD effects are unimportant), then the magnetic pressure will be strongly correlated with the gas density. When the thermal pressure is high, the gas density is low, and hence the magnetic pressure is low. In this regime, the momentum deposition rate from radiation pressure is governed by the recombination rate and hence scales with the square of the gas density. However, because the magnetic pressure is low, it typically has little effect on the density structure and hence little effect on the momentum deposition rate. Therefore, we expect the B-field to have only a negligible effect on the momentum absorbed by the warm \hii region. When the gas is cool, the gas density is higher and the magnetic pressure plays a far more important role in determining the structure of the shell. However, in this regime -- corresponding to the PDR and molecular gas regions -- the momentum deposition rate due to radiation pressure is determined by the total column density, which is unaffected by the strength of the magnetic pressure. Therefore, we also expect the B-field to have only a negligible effect on the momentum absorbed by the cooler gas.

Note that this does not mean that the B-field is of little consequence for the emission from the shell. In \cite{Pellegrini2007,Pellegrini2009}, we showed that the inclusion of a magnetic field in the EOS has a significant impact on the velocity structure of the line-emitting gas. Magnetic flux freezing can significantly reduce the PDR and molecular gas densities, decreasing the de-excitation rates of low critical density levels in $\rm S^+$, $\rm O^{++}$, $\rm N^+$, $\rm C^0$, $\rm C^+$ and  \twc. For example, the B-field may prevent C$^{+}$ from becoming collisionally de-excited when ionizing radiation pressure would otherwise drive gas densities high in the cold gas layer. 

We treat turbulent pressure support using a similar approximation. We assume that it does not significantly alter the dynamical evolution of the shell, but allow the user to account for small-scale turbulent pressure support in the shell. We do not include this effect by default because although molecular clouds are ubiquitously observed to be turbulent, the 
degree to which this turbulence acts to provide small-scale support is highly unclear \citep[see e.g.\ the discussion in][]{MacLow2004}.

Our RT and line emission calculations begin at the boundary of the wind blown bubble and the \hii region shell. In the absence of magnetic fields and turbulent support, we assume that the thermal pressure here is equal to the boundary pressure of the bubble $P_{\mathrm{b}}$:
\begin{equation} \label{press_equil_simple}
P_{\rm b} = P_{\rm therm}.
\end{equation}
However, when turbulence and a magnetic field are also included, we instead equate the boundary pressure with the total shell pressure:

\begin{equation} \label{press_equil}
P_{\mathrm{b}} = P_{\mathrm{therm}} + P_{\mathrm{B}} + P_{\mathrm{turb}}.
\end{equation}
The assumption of energy equipartition between the magnetic field and the turbulent field implies that  
\begin{equation}
P_{\mathrm{turb}} = P_{\mathrm{B}} = \frac{B^2}{8\pi},
\end{equation} and if we use a scaling law for the magnetic field in the form of 
\begin{equation}
B^2 = B_0^2\left( \frac{n}{n_0}\right)^{\gamma_{\mathrm{mag}} }
\end{equation}
as in \citet{Henney2005}, eq.~(\ref{press_equil}) becomes
\begin{equation}
P_{\mathrm{b}} = \frac{\mu_{\mathrm{n}}}{\mu_{\mathrm{p}}} n(R) k_B T_i + \frac{B_0^2}{4\pi}\left( \frac{n(R)}{n_0}\right)^{\gamma_{\mathrm{mag}}}
\end{equation}
with $B_0 = 48.7\,\mu$G, $n_0 = 116.1$\,\ccm, $\gamma_{\mathrm{mag}} = 4/3$, and $T_i = 10^4$\,K. In calculations with $B$-fields and turbulence we use this equation, which can be numerically solved for the ion number density at the inner edge of the shell $n(R)$, instead of Eq.~\ref{press_equil_simple}.

\section{Supplemental Material}
\label{appendix:SupPlots}

\begin{figure*}
    \centering
    \includegraphics[width=0.95\textwidth]{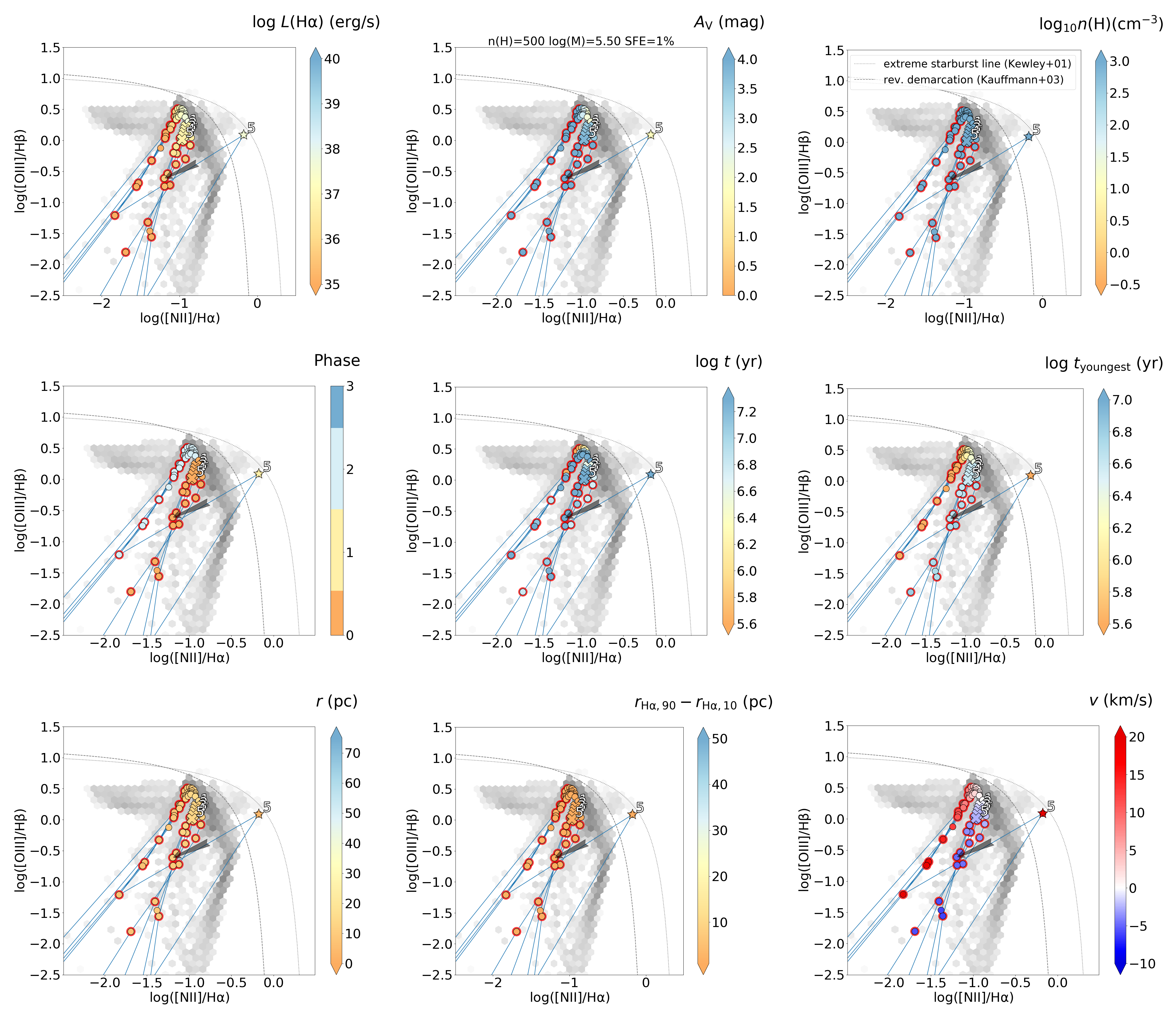}
    \caption{Same as Figure~\ref{fig:BPT_time_evol} but for a natal density of 500~${\rm cm^{-3}}$, cloud mass $\log_{10} M_{\rm cl} = 5.50$ and \SFE=1\%.}
    \label{fig:BTP_time_evol_high_n_low_SFE}
\end{figure*}

\begin{figure*}
    \centering
    \includegraphics[width=0.95\textwidth]{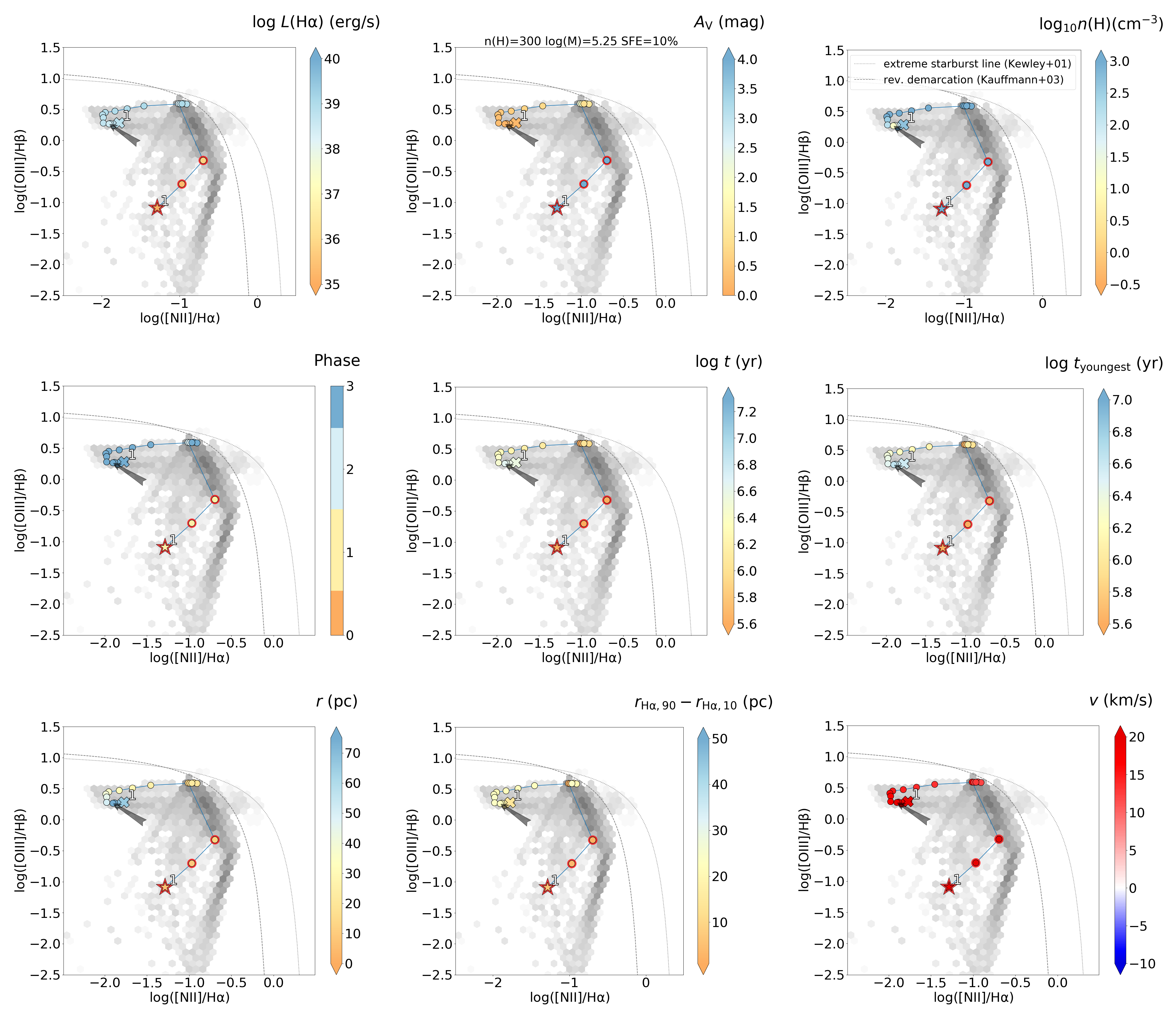}
    \caption{Same as Figure~\ref{fig:BPT_time_evol} but for a natal density of 300~${\rm cm^{-3}}$, cloud mass $\log_{10} M_{\rm cl} = 5.25$ and \SFE=10\%.}
    \label{fig:BTP_time_evol_med_n_high_SFE}
\end{figure*}

The following supplemental material provides some additional examples of evolutionary tracks along the lines of those shown in Figure~\ref{fig:BPT_time_evol}, but for different cloud and cluster properties. In Figure~\ref{fig:BTP_time_evol_high_n_low_SFE} we show an example of a low star formation efficiency cloud, with a high density. This results in many recollapses, as the cloud is not destroyed by feedback after many Myr. In Figure~\ref{fig:BTP_time_evol_med_n_high_SFE} we show a cloud with moderate density of 300~cm$^{-3}$, with a high 10\% SFE. Here feedback is sufficient to drive to cloud to large radii of almost 100~pc. The cloud becomes optically thin to ionizing radiation and the \NII/\Ha ratio decreases rapidly after shell expands beyond the natal cloud boundary. Finally, SNe mark a rapid decrease in pressure. While they do not destroy the cloud, which has already been disrupted by winds and radiation, the pressure loss associated with them leads to the density of the shell falling below that of the warm ionized medium. At this point, we consider the shell to have been destroyed.

\section{Default Emission Predictions}
\label{appendix:emp_default_lines}

The following is a list of the default set of line and continuum predictions made in {\sc warpfield-emp}. These include line and continuum emission, as well as flux-weighted opacities from dust and atomic and molecular processes. These can be calculated integrated across the nebula, or in projection. Note that this set does not include all of the lines that {\sc warpfield-emp} is capable of modelling; instead, it represents the subset that we expect to be most useful as tracers of \hii region, PDR or cloud properties.

\begin{table}
  \begin{center}
    \caption{Default {\sc warpfield-emp} Predicted Line List}
    \label{tab:lines}
    \begin{tabular}{l|c|c} 
     \textbf{Line} & \textbf{Wavelength} & \textbf{Region} \\
      \hline
        \Ha&6563$\,$\AA&\hii\\
        \Hb&4861$\,$\AA&\hii\\
        \NII&122$\,{\rm \micron}$&\hii\\
        \NII&205$\,{\rm \micron}$&\hii\\
        \NII&6584$\,$\AA&\hii\\
        \NII&5755$\,$\AA&\hii\\
        \OIII&88$\,{\rm \micron}$&\hii\\
        \OIII&5007$\,$\AA&\hii\\
        \OIII&4363$\,$\AA&\hii\\
        \OII&3727$\,$\AA&\hii\\
        \OII&3726$\,$\AA&\hii\\
        \OII&3729$\,$\AA&\hii\\
        \OII&7323$\,$\AA&\hii\\
        \OII&7332$\,$\AA&\hii\\
        \OI&63$\,{\rm \micron}$&\hii\\
        \OI&145$\,{\rm \micron}$&PDR\\
        \OI&6300$\,$\AA&PDR\\
        \SIV&10.5$\,{\rm \micron}$&\hii\\
        \SIII&18$\,{\rm \micron}$&\hii\\
        \SIII&33.5$\,{\rm \micron}$&\hii\\
        \SIII&9531$\,$\AA&\hii\\
        \SIII&9069$\,$\AA&\hii\\
        \SIII&6312$\,$\AA&\hii\\
        \SII&6716$\,$\AA&\hii\\
        \SII&6731$\,$\AA&\hii\\
        \SII&6720$\,$\AA&\hii\\
        \NeV&14$\,{\rm \micron}$&\hii\\
        \NeV&24$\,{\rm \micron}$&\hii\\
        \NeIII&15$\,{\rm \micron}$&\hii\\
        \NeIII&36$\,{\rm \micron}$&\hii\\
        \NeII&12.8$\,{\rm \micron}$&\hii\\
        \ArIII&9$\,{\rm \micron}$&\hii\\
        \ArIII&21.8$\,{\rm \micron}$&\hii\\
        \ArII&6.9$\,{\rm \micron}$&\hii\\
        \CII&158$\,{\rm \micron}$&PDR\\
        \CI&370$\,{\rm \micron}$&PDR\\
        \CI&609$\,{\rm \micron}$&PDR\\
        \twc&J(1-0)&Cloud\\
        \twc&J(2-1)&Cloud\\
        \twc&J(3-2)&Cloud\\
        \twc&J(4-3)&Cloud\\
        \twc&J(5-4)&Cloud\\
        \twc&J(6-5)&Cloud\\
        \twc&J(7-6)&Cloud\\
        \twc&J(8-7)&Cloud\\
        \twc&J(9-8)&Cloud\\
        \twc&J(10-9)&Cloud\\
        \twc&J(11-10)&Cloud\\
        \twc&J(12-11)&Cloud\\
        \twc&J(13-12)&Cloud\\
        HCO+&373.490$\,{\rm \micron}$&Cloud\\
        HCN&375.844$\,{\rm \micron}$&Cloud\\
        \SiII&34$\,{\rm \micron}$&HII/PDR\\
        H$_2$&2.121$\,{\rm \micron}$&PDR\\
        H$_2$ S(0)&5.50$\,{\rm \micron}$&PDR/Cloud\\
        H$_2$ S(1)&6.1$\,{\rm \micron}$&PDR/Cloud\\
        H$_2$ S(2)&6.9$\,{\rm \micron}$&PDR/Cloud\\
        H$_2$ S(3)&8.0$\,{\rm \micron}$&PDR/Cloud\\
        H$_2$ S(4)&9.6$\,{\rm \micron}$&PDR/Cloud\\
        H$_2$ S(5)&12.2$\,{\rm \micron}$&PDR/Cloud\\
        H$_2$ S(6)&17.0$\,{\rm \micron}$&PDR/Cloud\\
        H$_2$ S(7)&28.2$\,{\rm \micron}$&PDR/Cloud\\
    \end{tabular}
  \end{center}
\end{table}

\newpage

\begin{table}
  \begin{center}
    \caption{Default {\sc warpfield-emp} Predicted Continuum List}
    \label{tab:continuum}
    \begin{tabular}{l|c|c|c} 
     \textbf{Phot. Band} & \textbf{Wavelength} & \textbf{Region} \\
      \hline
        PAC3&160$\,{\rm \micron}$&HII+PDF+CLOUD\\
        PAC2&100$\,{\rm \micron}$&HII+PDF+CLOUD\\
        PAC1&70$\,{\rm \micron}$&HII+PDF+CLOUD\\
        SPR3&500$\,{\rm \micron}$&HII+PDF+CLOUD\\
        SPR2&350$\,{\rm \micron}$&HII+PDF+CLOUD\\
        SPR1&250$\,{\rm \micron}$&HII+PDF+CLOUD\\
        IRAC&8$\,{\rm \micron}$&HII+PDF+CLOUD\\
        IRAC&5.8$\,{\rm \micron}$&HII+PDF+CLOUD\\
        IRAC&4.5$\,{\rm \micron}$&HII+PDF+CLOUD\\
        IRAC&3.6$\,{\rm \micron}$&HII+PDF+CLOUD\\
        MIPS&160$\,{\rm \micron}$&HII+PDF+CLOUD\\
        MIPS&70$\,{\rm \micron}$&HII+PDF+CLOUD\\
        MIPS&24$\,{\rm \micron}$&HII+PDF+CLOUD\\
        F100&100$\,{\rm \micron}$&HII+PDF+CLOUD\\
        F60&60$\,{\rm \micron}$&HII+PDF+CLOUD\\
        F25&25$\,{\rm \micron}$&HII+PDF+CLOUD\\
        F12&12$\,{\rm \micron}$&HII+PDF+CLOUD\\
        TIR&3-1100$\,{\rm \micron}$&HII+PDF+CLOUD\\
        \hline
    \end{tabular}
  \end{center}
     \begin{small}
      The photometric bands listed here correspond to different observatories: PAC and SPR refer to Herschel's PACS and SPIRE instruments. The IRAC and MIPS bands refer to Spitzer instruments and filter band-passes, and F100, 60, 25 and 12 are generic bands which have been added to {\sc Cloudy}.
\end{small}
\end{table}

%
\bibliographystyle{mn2e}  
\bibliography{library}

\end{document}